\pdfoutput=1
\RequirePackage{ifpdf}
\ifpdf 
\documentclass[pdftex]{sigma}
\else
\documentclass{sigma}
\fi

\usepackage[all]{xy}

{\theoremstyle{definition}
\newtheorem*{notation}{Notation}}

\newtheorem{Theorem}{Theorem}[section]
\newtheorem{Corollary}[Theorem]{Corollary}
\newtheorem{Lemma}[Theorem]{Lemma}
\newtheorem{Proposition}[Theorem]{Proposition}
\newtheorem{Observation}[Theorem]{Observation}
{\theoremstyle{definition}
\newtheorem{Definition}[Theorem]{Definition}
\newtheorem{Remark}[Theorem]{Remark}
}

\numberwithin{equation}{section}

\DeclareMathOperator{\Ker}{Ker}
\DeclareMathOperator{\card}{card}

\begin{document}

\allowdisplaybreaks

\renewcommand{\PaperNumber}{062}

\FirstPageHeading

\ShortArticleName{Period Matrices of Real Riemann Surfaces and Fundamental Domains}

\ArticleName{Period Matrices of Real Riemann Surfaces\\
and Fundamental Domains}

\Author{Pietro GIAVEDONI}

\AuthorNameForHeading{P.~Giavedoni}

\Address{Faculty of Mathematics, University of Vienna, Oskar-Morgenstern-Platz 1, 1090 Wien, Austria}
\Email{\href{mailto:pietro.giavedoni@univie.ac.at}{pietro.giavedoni@univie.ac.at},
\href{mailto:addenaro@gmail.com}{addenaro@gmail.com}}

\ArticleDates{Received March 01, 2013, in f\/inal form October 14, 2013; Published online October 22, 2013}

\Abstract{For some positive integers $g$ and $n$ we consider a~subgroup $\mathbb{G}_{g,n}$ of the
$2g$-dimensional modular group keeping invariant a~certain locus $\mathcal{W}_{g,n}$ in the Siegel upper
half plane of degree $g$.
We address the problem of describing a~fundamental domain for the modular action of the subgroup on
$\mathcal{W}_{g,n}$.
Our motivation comes from geometry: $g$ and $n$ represent the genus and the number of ovals of a~generic
real Riemann surface of separated type; the locus $\mathcal{W}_{g,n}$ contains the corresponding period
matrix computed with respect to some specif\/ic basis in the homology.
In this paper we formulate a~general procedure to solve the problem when $g$ is even and $n$ equals one.
For $g$ equal to two or four the explicit calculations are worked out in full detail.}


\Classification{14P05; 57S30; 11F46}

\section{Introduction}

The \textit{Siegel upper half plane of degree $g$}, usually denoted $\mathbb{H}_{g}$, is def\/ined as the
space of $g\times g$ complex, symmetric matrices whose imaginary part is positive def\/inite.
The \textit{modular group of dimension $2g$} is the group of $2g\times 2g$ symplectic matrices whose
entries are integer numbers.
We will denote it by the symbol $\mathbf{Sp}(2g,\mathbb{Z})$.

There exists a~well-known action of the modular group on the Siegel upper half plane, the so called
\textit{modular action}:
\begin{gather*}
\mathfrak{M}:\ \mathbf{Sp}(2g,\mathbb{Z})\times\mathbb{H}_{g}\longrightarrow\mathbb{H}_{g}.
\end{gather*}
To every matrix
\begin{gather*}
G=\left(\begin{matrix}P&Q\\R&S\end{matrix}\right)\in\mathbf{Sp}(2g,\mathbb{Z})
\end{gather*}
and every point $\mathbf{w}$ of $\mathbb{H}_{g}$ it associates
\begin{gather*}
\mathfrak{M}(G,\mathbf{w})=(P\mathbf{w}+Q)(R\mathbf{w}+S)^{-1}.
\end{gather*}
Now, let $g$ and $n$ be positive integers such that
$
g+1=2p+n
$
for some non-negative integer~$p$.
Let us def\/ine the space
\begin{gather*}
\mathcal{W}_{g,n}=\big\{\mathbf{w}\in\mathbb{H}_{g}
\;
\text{such that}
\;
V\mathbf{w}V=-\overline{\mathbf{w}}
\big\},
\end{gather*}
where the matrix $V$ is given by
\begin{gather*}
V=\left(
\begin{matrix}0&{\rm Id}_{p}&0
\\
{\rm Id}_{p}&0&0
\\
0&0&{\rm Id}_{n-1}
\end{matrix}
\right).
\end{gather*}
Let us also introduce the group
\begin{gather*}
\mathbb{G}_{g,n}=\big\{G\in\mathbf{Sp}(2g,\mathbb{Z})
\;
\text{such that}
\;
GT=TG\big\},
\end{gather*}
where $T$ denotes the $2g\times 2g$ matrix
\begin{gather*}
T=\left(\begin{matrix}V&0\\0&-V\end{matrix}\right).
\end{gather*}
The modular action can be restricted to an action of $\mathbb{G}_{g,n}$ on the space $\mathcal{W}_{g,n}$:
for every matrix
\begin{gather*}
G=\left(\begin{matrix}P&Q\\R&S\end{matrix}\right)\in\mathbb{G}_{g,n}
\end{gather*}
and every point
$
\mathbf{w}\in\mathcal{W}_{g,n}
$
one has that
$
\mathfrak{M}(G,\mathbf{w})\in\mathcal{W}_{g,n}$.
Indeed, from the def\/inition of $\mathbb{G}_{g,n}$ one can immediately deduce the following relations:
\begin{gather*}
P=VPV,
\qquad
Q=-VQV,
\qquad
R=-VRV,
\qquad
S=VSV.
\end{gather*}
These ones yield
\begin{gather*}
V [\mathfrak{M}(G,\mathbf{w}) ]V=V(P\mathbf{w}+Q)(R\mathbf{w}+S)^{-1}
\\
\hphantom{V [\mathfrak{M}(G,\mathbf{w}) ]V}{}
=\big[(VPV) (V\mathbf{w}
V )+ (VQV )\big]\big[ (VRV )(V\mathbf{w}V)+(VSV)\big]^{-1}
\\
\hphantom{V [\mathfrak{M}(G,\mathbf{w}) ]V}{}
=-\big[P\overline{\mathbf{w}}+Q\big]\big[R\overline{\mathbf{w}}+S\big]^{-1}
=-\overline{\mathfrak{M}(G,\mathbf{w})}.
\end{gather*}

We wish to address the problem of a~fundamental domain for the modular action of the group
$\mathbb{G}_{g,n}\subset \mathbf{Sp}(2g,\mathbb{Z})$ on the space $\mathcal{W}_{g,n}$.
Its def\/inition is recalled here below:
\begin{Definition}\label{def_dom_fond}
We say that two points $\mathbf{w}_1$ and $\mathbf{w}_2$ of $\mathcal{W}_{g,n}$ are equivalent if there
exists $G\in\mathbb{G}_{g,n}$ such that
$
\mathbf{w}_{1}=\mathfrak{M} (G,\mathbf{w}_{2} )$.
A fundamental domain for $\mathfrak{M}$ is a~closed subset $\mathcal{D}$ of $\mathcal{W}_{g,n}$ whose
interior part is connected, which satisf\/ies the following three properties:
\begin{itemize}\itemsep=0pt
\item[i)] For every $\mathbf{w}\in\mathcal{W}_{g,n}$ there exists an equivalent point $\tilde{\mathbf{w}}$
belonging to $\mathcal{D}$.
\item[ii)] If two distinct points, $\mathbf{w}$ and $\tilde{\mathbf{w}}$, belonging to $\mathcal{D}$, are
equivalent, then they both belong to the boundary of $\mathcal{D}$.
\item[iii)] Every set of equivalent points contained in $\mathcal{D}$ has a~f\/inite number of elements.
\end{itemize}
\end{Definition}
The case in which $n$ equals $g+1$ is easily reconducted to the reduction theory of positive def\/inite
quadratic forms and was already studied by several mathematicians from the last century.
(The f\/irst achievements in a~general treatment of this topic are due to Minkowski~\cite{Minkowski};
see~\cite{Ryshkov} for an account on more recent results.)

In this article we solve the problem when $g = 2g_0$ is an even number, and $n$ equals one.
In this case, the space $\mathcal{W}_{2g_0,1}$ has the following simple characterization:
\begin{gather*}
\mathcal{W}_{2g_0,1}=\left\{\mathbf{w}\in\mathbb{H}_{g}
\;
\text{such that}
\;
\mathbf{w}=\left[\begin{matrix}\mathbf{z}&\mathbf{x}\\-\overline{\mathbf{x}}&-\overline{\mathbf{z}}\end{matrix}\right];
\;
\mathbf{z},\mathbf{x}\in {\rm Mat} (g_0,\mathbb{C} )\right\}.
\end{gather*}
For the group $\mathbb{G}_{2g_0,1}$, instead, one has{\samepage
\begin{gather*}
\mathbb{G}_{2g_0,1}=\left\{G\in\mathbf{Sp} (4g_0,\mathbb{Z} )
\;
\text{such that}
\;
G=\left[
\begin{matrix}
A&B&C&D
\\
B&A&-D&-C
\\
E&F&G&H
\\
-F&-E&H&G
\end{matrix}
\right]\right\}.
\end{gather*}
It is understood that the matrices $A$, $B$, $C$, $D$, $E$, $F$, $G$ and $H$ belong to ${\rm Mat}(g_0,\mathbb{Z})$.}

\begin{figure}[t]
\centering
\includegraphics[scale=1]{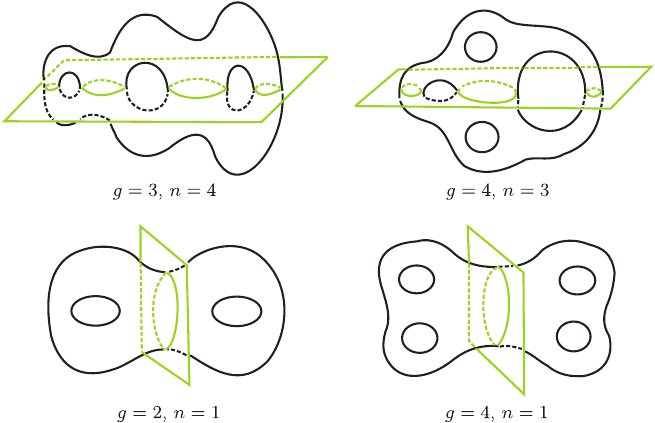}

\caption{Examples of \textit{separated} real Riemann surfaces with dif\/ferent genus~$g$ and number of
ovals~$n$.
The anti-holomorphic involution~$r$ can be thought of as the ref\/lection with respect to the green plane.
The ovals are also drawn in green.}
\label{Miracolo}
\end{figure}

Our main motivation in considering this problem comes from the theory of \textit{real Riemann surfaces}
(see~\cite{Cirre,Fay} and~\cite{NatMod} for a~complete account of this subject).
A compact Riemann surface~$\Gamma$ of genus~$g$ is said to be \textit{real} or \textit{symmetric} when it
is endowed with an anti-holomorphic involution~$r$.
One can consider the locus of points which are invariant with respect to~$r$.
Its connected components are at most $g+1$ and they are usually called the \textit{ovals} of the surface.
When~$\Gamma$ with this locus removed has two connected components, it is said to be \textit{separated}
(see Fig.~\ref{Miracolo}).
Let this be the case, and assume that the number of ovals is~$n$.
Then it is possible to f\/ix a~basis in the homology of the form\footnote{In particular, this implies that
the genus and the number of ovals of a~separated, real Riemann surface need to satisfy the following
relation:
$n\equiv g+1$ ${\rm mod}\; 2$.
For more details see~\cite[Theorem~1.5.3]{Cirre}.}
\begin{gather}
a_1^{\prime},\quad a_2^{\prime},\quad \ldots,\quad a_p^{\prime},\quad
a_1^{\prime\prime},\quad a_2^{\prime\prime},\quad \ldots,\quad
a_p^{\prime\prime},\quad a_1,\quad a_2,\quad \ldots,\quad a_{n-1},
\nonumber
\\
b_1^{\prime},\quad b_2^{\prime},\quad \ldots,\quad b_p^{\prime},\quad
b_1^{\prime\prime},\quad b_2^{\prime\prime},\quad \ldots,\quad
b_p^{\prime\prime},\quad b_1,\quad b_2,\quad \ldots,\quad b_{n-1}
\label{Intro_1}
\end{gather}
satisfying the following conditions:
\begin{gather}\label{Intro_2}
a_{k}^{\prime}\circ b_{k}^{\prime}=a_{k}^{\prime\prime}\circ b_{k}^{\prime\prime}=a_{j}\circ b_{j}=1
\end{gather}
for every $k=1,2,\ldots,p$ and every $j=1,2,\ldots,n-1$, the intersection number of any two other elements
of the basis being zero;
\begin{gather}
r_{\star}\left(a_k^{\prime}\right)=-a_k^{\prime\prime},
\qquad
k=1,2,\ldots,p,
\nonumber
\\
r_{\star}\left(b_k^{\prime}\right)=b_k^{\prime\prime},
\qquad
k=1,2,\ldots,p,
\nonumber
\\
r_{\star} (a_j )=-a_j,
\qquad
j=1,2,\ldots,n-1,
\nonumber
\\
r_{\star} (b_j )=b_j,
\qquad
j=1,2,\ldots,n-1,\label{Intro_3}
\end{gather}
where $r_{\star}$ denotes the morphism of the homology into itself induced by~$r$.

Given any Riemann surface (not necessarily a~real one) with a~f\/ixed basis in the homology $\mathcal{B}$,
one can compute the corresponding \textit{period matrix}\footnote{In this procedure, the normalized
holomorphic dif\/ferentials
$
\omega_1^{\prime}$,
$
\omega_2^{\prime}$,
$
\ldots$,
$
\omega_p^{\prime}$,
$
\omega_1^{\prime\prime}$,
$
\omega_2^{\prime\prime}$,
$
\ldots$,
$
\omega_p^{\prime\prime}$,
$
\omega_1$,
$
\omega_2$,
$
\ldots$,
$
\omega_{n-1}$
are involved.
Here we have been assuming the convention
$
\oint_{a_k^{\prime}}\omega_k^{\prime}=\oint_{a_k^{\prime\prime}}\omega_k^{\prime\prime}=\oint_{a_j}\omega_j=1
$
for $k = 1,2,\ldots,p$ and $j = 1,2,\ldots,n-1$.}.
This is a~point in the Siegel upper half plane (see~\cite{SiegelII} and~\cite{Fay}).
The issue of determining the locus of period matrices in $\mathbb{H}_g$ is the famous Schottky problem and
it is deeply connected with the KP equation~\cite{BeaPro,DebSch,GruSch}.
Any other basis in the homology $\tilde{\mathcal{B}}$ on the same surface can be expressed in terms of the
cycles of $\mathcal{B}$ by means of a~modular matrix.
If this is blockwise written as
\begin{gather*}
\left(\begin{matrix}P&Q\\R&S\end{matrix}\right)\in\mathbf{Sp}(2g,\mathbb{Z})
\end{gather*}
then the corresponding period matrices $\mathbf{t}$ and $\tilde{\mathbf{t}}$ are related by the modular
transformation
\begin{gather*}
\tilde{\mathbf{t}}= (P\mathbf{t}+Q ) (R\mathbf{t}+S )^{-1}.
\end{gather*}
Now, in the case of a~real Riemann surface with a~basis $\mathcal{B}$ of the
form~\eqref{Intro_1}--\eqref{Intro_3}, the period matrix belongs to the space $\mathcal{W}_{g,n}$.
Moreover, the changes of basis in the homology which preserve the form~\eqref{Intro_1}--\eqref{Intro_3} are
exactly the ones given by matrices of $\mathbb{G}_{g,n}$.

Bases in the homology of the form~\eqref{Intro_1}--\eqref{Intro_3} can also be found on real Riemann surfaces
without f\/ix points for $r$.
This can be deduced, for example, from~\cite[Lemma~1]{CosPoi}.
In this case~$n$ is just a~formal parameter, equal to 1 or 2 if $g$ is even or odd respectively.
In view of this fact, also the problem of a~fundamental domain for period matrices of such real Riemann
surfaces can be formulated in terms of the modular action of the corresponding group
$\mathbb{G}_{g,n}\subset \mathbf{Sp}(2g,\mathbb{Z})$ on the space $\mathcal{W}_{g,n}$.

Real Riemann surfaces were extensively applied in the theory of integrable systems (see~\cite{Matveev}).
In~\cite{Dubrovin-Segur}, the solution of our problem in one particular case was successfully applied to
the study of the parameter space of algebro-geometric solutions of the KP-2 equation.

As we have already said, in this paper we formulate a~general procedure to determine a~fundamental domain
for the modular action of the group $\mathbb{G}_{2g_0,1}$ on the space $\mathcal{W}_{2g_0,1}$.
We brief\/ly sketch our methods and results here below:

Our f\/irst step consists in a~reformulation of the problem (Section~\ref{sez:Reformulation}).
Let us denote with $\mathbf{GL}(g,\mathbb{Z})$ the group of all $g$-dimensional, unimodular
matrices with integer entries.
Let us also introduce the space ${\rm Sym}_{>0}(g,\mathbb{R})$ of all real symmetric and
positive def\/inite matrices of di\-men\-sion~$g$.
The \textit{congruence action}
\begin{gather}
\label{def_cong_1}
\mathfrak{C}:\ {\rm Sym}_{>0}(g,\mathbb{R})\times\mathbf{GL} (g,\mathbb{Z}
 )\longrightarrow{\rm Sym}_{>0}(g,\mathbb{R})
\end{gather}
is def\/ined as follows
\begin{gather}
\label{def_cong_2}
\mathfrak{C} (\Sigma,G )=G^{T}\Sigma G.
\end{gather}
The modular and the congruence actions are closely related by the following result: for every
\begin{gather*}
\mathbf{w}=\lambda+\imath\mu
\;
\in\mathbb{H}_{g}
\end{gather*}
let us def\/ine
\begin{gather*}
\mathbf{F} (\mathbf{w} )=\left[
\begin{matrix}\mu^{-1}&-\mu^{-1}\lambda
\\
-\lambda\mu^{-1}&\mu+\lambda\mu^{-1}\lambda
\end{matrix}
\right].
\end{gather*}

\noindent
{\bf Theorem 2.1 (Siegel).} {\it  The map $\mathbf{F}$ is smooth and injective from $\mathbb{H}_g$ into ${\rm
Sym}_{>0}\left( 2g,\mathbb{R}\right)$.
Moreover, the diagram below is commutative for every $G$ belonging to $\mathbb{G}_{2g_0,1}$:}
\begin{gather*}
\begin{gathered}\xymatrix{\mathbb{H}_{g}\ar[r]^{\mathfrak{M}_{G}}\ar[d]^{\mathbf{F}}&\mathbb{H}_{g}\ar[d]^{\mathbf{F}}
\\
{\rm Sym}_{>0}(2g,\mathbb{R})\ar[r]^{\mathfrak{C}_{G^{-1}}}&{\rm Sym}_{>0}\left(2g,\mathbb{R}
\right)}
\end{gathered}
\qquad
G\in\mathbf{Sp}(2g,\mathbb{Z}).
\end{gather*}

Here we have introduced the notation:
\begin{gather*}
\mathfrak{M}_{G}:\ \mathbb{H}_{g}\longrightarrow\mathbb{H}_{g}, \qquad \mathfrak{M}_{G}(\mathbf{w})=\mathfrak{M}(G,\mathbf{w}),
\\
\mathfrak{C}_{G}:\ {\rm Sym}_{>0}(n,\mathbb{R})\longrightarrow{\rm Sym}_{>0}(n,\mathbb{R}),
\qquad
\mathfrak{C}_{G}(\Sigma)=\mathfrak{C}(\Sigma,G).
\end{gather*}
In view of our purposes, we consider the restriction of the map $\mathbf{F}$ to $\mathcal{W}_{2g_0,1}$.
The image of this last one is the space
\begin{gather*}
\mathcal{S}_{2g_0,1}:=\left\{\left[
\begin{matrix}\alpha&\beta&\gamma&\delta
\\
\beta&\alpha&-\delta&-\gamma
\\
\gamma^T&-\delta^T&\xi&\eta
\\
\delta^T&-\gamma^T&\eta&\xi
\\
\end{matrix}
\right]
\in
\mathbf{Sp}(4g_0,\mathbb{R})\cap{\rm Sym}_{>0}(4g_0,\mathbb{R})\right\}.
\end{gather*}
The congruence action can be restricted to an action of the group $\mathbb{G}_{2g_0,1}$ on
$\mathcal{S}_{2g_0,1}$.
Moreover one has the following

\medskip

\noindent
{\bf Corollary 2.2.} {\it Let $\mathcal{D}^{\prime}$ a~fundamental domain for the congruence action of the group
$\mathbb{G}_{2g_0,1}$ on~$\mathcal{S}_{2g_0,1}$.
Then $
\mathcal{D}=\mathbf{F}^{-1}\left(\mathcal{\mathcal{D}^{\prime}}\right)$
is a~fundamental domain for the modular action of the same~$\mathbb{G}_{2g_0,1}$ on~$\mathcal{W}_{2g_0,1}$.}

\medskip

As a~consequence, the original issue is equivalent to the quest of such a~$\mathcal{D}^{\prime}$.
(See~\cite{Riera} for an interpretation of this reformulation in terms of Klein triples.)
The main technical tool to tackle this new problem is the map $\mathbf{P}$, introduced in
Section~\ref{sez:ReductionToolkit}.
Due to its several remarkable properties this might turn out to be of general interest by itself, beyond
its role in this specif\/ic context.
To the best of our knowledge, this map was never considered in the literature before.

The domain of $\mathbf{P}$ is the group $\mathbb{G}_{2g_0,1}(\mathbb{R})$ of all
$4g_0$-dimensional, real and symplectic matrices of the form
\begin{gather*}
\left[
\begin{matrix}\alpha&\beta&\gamma&\delta
\\
\beta&\alpha&-\delta&-\gamma
\\
\pi&\rho&\xi&\eta
\\
-\rho&-\pi&\eta&\xi
\end{matrix}
\right]\in\mathbf{Sp}(4g_0,\mathbb{R}).
\end{gather*}
Both $\mathcal{S}_{2g_0, 1}$ and $\mathbb{G}_{2g_0,1}$ are contained in it.
Its explicit def\/inition is very simple:
\begin{gather*}
\mathbf{P}\left(\left[
\begin{matrix}
\alpha&\beta&\gamma&\delta
\\
\beta&\alpha&-\delta&-\gamma
\\
\pi&\rho&\xi&\eta
\\
-\rho&-\pi&\eta&\xi
\end{matrix}
\right]\right)=\left[
\begin{matrix}
\alpha+\beta&\gamma-\delta
\\
\pi+\rho&\xi-\eta
\end{matrix}
\right].
\end{gather*}
After this, $\mathbf{P}$ turns out to be a~bijection onto $\mathbf{GL}(2g_0,\mathbb{R})$, the
space of all real and invertible matrices of dimension $2g_0$.
It maps $\mathcal{S}_{2g_0,1}$ onto ${\rm Sym}_{>0}(2g_0,\mathbb{R})$.
Moreover $\mathbf{P}$ respects both the matrix product and transposition.

As a~consequence, one has the following

\medskip

\noindent
{\bf Proposition 4.1.} {\it The diagram below is commutative for every $G$ belonging to $\mathbb{G}_{2g_0,1}$:}
\begin{gather*}
\begin{gathered}
\xymatrix{\mathcal{S}_{2g_0,1}\ar[r]^{\mathfrak{C}_{G}}\ar[d]^{\mathbf{P}}&\mathcal{S}_{2g_0,1}
\ar[d]^{\mathbf{P}}
\\
{\rm Sym}_{>0}(2g_0,\mathbb{R})\ar[r]^{\mathfrak{C}_{\mathbf{P}(G)}}&{\rm Sym}_{>0}
(2g_0,\mathbb{R})}
\end{gathered}
\qquad
G\in\mathbb{G}_{2g_0,1}.
\end{gather*}

This construction allows a~further reformulation of the problem, in the same spirit as above.
Let us denote by $J$ the \emph{standard symplectic matrix}:
\begin{gather}\label{definizione_simplettica_standard}
J=\left[\begin{matrix}0&{\rm Id}\\-{\rm Id}&0\end{matrix}\right].
\end{gather}
Its dimension will be clear from the context.
Let us introduce the group
\begin{gather*}
\mathbb{K}_{2g_0,1}:=\big\{h\in\mathbf{GL}(2g_0,\mathbb{Z})
\;
\text{such that}
\;
h^{T}Jh\equiv J
\ {\rm mod}\; 2\big\},
\end{gather*}
which is a~f\/inite-index subgroup of $\mathbf{GL}(2g_0,\mathbb{Z})$.
This is actually the image of the group $\mathbb{G}_{2g_0,1}$ via the map $\mathbf{P}$.

\medskip

\noindent
{\bf Theorem 4.2.} {\it Let $\mathcal{D}^{\prime\prime}$ be a~fundamental domain for the congruence action of
$\,\mathbb{K}_{2g_0,1}$ on $\mathcal{S}_{2g_0,1}$.
Then
$
\mathcal{D}^{\prime}=\mathbf{P}^{-1}\left(\mathcal{D}^{\prime\prime}\right)
$
is a~fundamental domain for the congruence action of $\mathbb{G}_{2g_0,1}$ on $\mathcal{S}_{2g_0,1}$.}

\medskip

The original problem is then equivalent to the quest of such a~$\mathcal{D}^{\prime\prime}$.
This second reformulation appears in Section~\ref{sez:Reduction}.
It is given the name of reduction of the problem because it halves, so to say, the dimension of the
matrices involved in the issue.
The advantage of these subsequent reformulations is that we f\/inally get to an explicitly solvable problem.
Indeed, the Minkowski reduction theory provides a~fundamental domain for the congruence action of the whole
group $\mathbf{GL}(2g_0,\mathbb{Z})$ on ${\rm Sym}_{>0}(2g_0,\mathbb{R})$.
On the other side, the index of $\mathbb{K}_{2g_0,1}$ in $\mathbf{GL}(2g_0,\mathbb{Z})$ is
f\/inite.
In view of these two facts, a~fundamental domain $\mathcal{D}^{\prime\prime}$ for the congruence action of
$\mathbb{K}_{2g_0,1}$ on ${\rm Sym}_{>0}(2g_0,\mathbb{R})$ can be computed by means of standard
``gluing'' techniques.

The case in which $2g_0$ equals 2 is particularly interesting (Section~\ref{sez:CaseGenusTwo}).
It exhibits the remarkab\-le peculiarity that the group $\mathbb{K}_{2,1}$ coincides with the whole
$\mathbf{GL}(2,\mathbb{Z})$.
As a~f\/irst consequence, the two reformulations above reconduct the original problem to a~classical one
already solved: the reduction of positive def\/inite binary quadratic forms.
No usage of ``gluing'' techniques is required in this case.
By a~straightforward calculation one can reobtain a~result already known to Silhol \cite{SilAlg}. Let us
denote by
\begin{gather*}
\mathbf{w}=\left(\begin{matrix}\gamma+\imath\delta&\imath\beta\\\imath\beta&-\gamma+\imath\delta\end{matrix}\right),
\qquad
\beta,\gamma,\delta\in\mathbb{R},
\end{gather*}
the generic element of $\mathcal{W}_{2,1}$.

\medskip

\noindent
{\bf Theorem 5.3.} {\it A fundamental domain $\mathcal{D}$ for the modular action $\mathfrak{M}$ of the group
$\mathbb{G}_{2,1}\subset\mathbf{Sp}(4,\mathbb{Z})$ on this last one is given by the following
system of inequalities:
\begin{gather*}
\mathcal{D}:
\
1\leq\gamma^2+\delta^2-\beta^2,
\qquad
0\leq\gamma\leq\frac{1}{2},
\qquad
\delta>|\beta|.
\end{gather*}
$($Notice that the third line of this system just guarantees that the imaginary part of $\mathbf{w}$ is
positive def\/inite.$)$}

\medskip

Moreover, a~deeper understanding of the mathematical structure is possible in this case.
Let $G$ belong to $\mathbb{G}_{2,1}$.
One has that either
\begin{gather}\label{primo_tipo}
G=\left[
\begin{matrix}
a&0&c&0
\\
0&a&0&-c
\\
e&0&g&0
\\
0&-e&0&g
\end{matrix}
\right],
\qquad
ag-ec=1
\end{gather}
or
\begin{gather}\label{secondo_tipo}
G=\left[
\begin{matrix}
0&b&0&d
\\
b&0&-d&0
\\
0&f&0&h
\\
-f&0&h&0
\end{matrix}
\right],
\qquad
bh-fd=1.
\end{gather}

\noindent
{\bf Theorem 5.4.} {\it Let us introduce the following new system of coordinates on $\mathcal{W}_{2,1}$:
\begin{gather*}
\mathcal{I}(\mathbf{w})=\frac{\delta-\beta}{\delta+\beta},
\qquad
\boldsymbol{\tau}(\mathbf{w})=\gamma+\imath\sqrt{\delta^2-\beta^2},
\qquad
\mathbf{w}\in\mathcal{W}_{2,1}.
\end{gather*}
Now consider a~modular transformation
\begin{gather*}
\tilde{\mathbf{w}}=\mathfrak{M}(G,\mathbf{w}),
\qquad
\mathbf{w}\in\mathcal{W}_{2,1},
\qquad
G\in\mathbb{G}_{2,1}.
\end{gather*}
In the new coordinates it acts as follows:
\begin{gather*}
\mathcal{I}(\tilde{\mathbf{w}})=\mathcal{I}(\mathbf{w})
\end{gather*}
and
\begin{gather*}
\boldsymbol{\tau}(\tilde{\mathbf{w}})
=\frac{a\boldsymbol{\tau}(\mathbf{w})+c}{e\boldsymbol{\tau}(\mathbf{w})+g}
\end{gather*}
if $G$ has the form~\eqref{primo_tipo}, or
\begin{gather*}
\boldsymbol{\tau}(\tilde{\mathbf{w}})
=\frac{b\overline{\boldsymbol{\tau}(\mathbf{w})}-d}{f\overline{\boldsymbol{\tau}(\mathbf{w})}-h}
\end{gather*}
if~\eqref{secondo_tipo} holds, instead.}

\medskip

In other words, we f\/ind out that $\mathcal{I}$ is an invariant quantity: it is not af\/fected by the
modular action of $\mathbb{G}_{2,1}$, which concentrates only on the second coordinate $\boldsymbol{\tau}$.
This last one just undergoes a~Moebius transformation, in some cases composed with a~complex conjugation.
This fact has some interesting geometrical consequences.
To every genus two, real Riemann surface $(\Gamma,r)$ of separated type with just one oval,
or with no ovals at all, one can associate a~couple
\begin{gather*}
(\Gamma,r)\longrightarrow (\mathcal{I},\boldsymbol{\Delta}),
\end{gather*}
where $\mathcal{I}\in \mathbb{R}$ is the quantity introduced above and $\boldsymbol{\Delta}$ is a~genus one
Riemann surface determined by $\boldsymbol{\tau}(\mathbf{w})$.
A satisfactory geometrical interpretation of this correspondence, together with its generalization to
higher genera is the goal of current investigations.

The case $g\geq 4$ is less immediate.
In section 6 we formulate a~procedure to determine $\mathcal{D}^{\prime\prime}$ using the results of
Minkowski reduction theory.
For the case~$g$ equal to~4 we obtain the following, very explicit result:

\medskip

\noindent
{\bf Theorem 6.9.} {\it Let $\mathcal{M}$ be the Minkowski fundamental domain for the congruence action of
$\mathbf{GL}(4,\mathbb{Z})$ on ${\rm Sym}_{>0}(4,\mathbb{R})$.
A fundamental domain $\mathcal{D}^{\prime\prime}$ for the congruence action of $\mathbb{K}_{4,1}$ on the
same space is given by
\begin{gather*}
\mathcal{D}^{\prime\prime}=\bigcup_{j=1}^{28}\mathfrak{C}(\mathcal{M},g_j).
\end{gather*}
The explicit expressions for $g_1,g_2,\ldots,g_{28}\in \mathbf{GL}(4,\mathbb{Z})$ are given
in~\eqref{fondamentali}, \eqref{tutti_definitivi}.}

\medskip

Since completing this paper, we have noted that some of the results of
Section~\ref{sez:CaseGenusTwo} are closely related to the ones presented in~\cite{SilAlg,Silhol}.
This connection is discussed at the end of Section~\ref{sez:GeneralCase}.

\section{Reformulation of the problem}
\label{sez:Reformulation}

In this section we show how the original problem can be reformulated in terms of a~restriction of the
congruence action $\mathfrak{C}$, introduced in~\eqref{def_cong_1},~\eqref{def_cong_2}.
For every matrix
$
\mathbf{w}=\lambda+\imath\mu
\in\mathbb{H}_{g}
$
belonging to the Siegel upper half plane of degree $g$, let us def\/ine
\begin{gather}\label{def_Sigma}
\mathbf{F}(\mathbf{w})=\left[
\begin{matrix}
\mu^{-1}&-\mu^{-1}\lambda
\\
-\lambda\mu^{-1}&\mu+\lambda\mu^{-1}\lambda
\end{matrix}
\right],
\qquad
\mathbf{w}\in\mathbb{H}_{g}.
\end{gather}
This map already appeared in Siegel's investigations on the modular group, from which we drew some
inspiration.
It relates the modular and the congruence actions:
\begin{Theorem}\label{teo_Siegel}\qquad
\begin{itemize} \itemsep=0pt
\item[$i)$] $\mathbf{F}$ is a~bijection onto the space of all real, symmetric and positive definite
matrices of dimension $2g$ which are also symplectic.
\item[$ii)$] For every $G$ belonging to the modular group, the following diagram is commutative:
\begin{gather*}
\begin{gathered}
\xymatrix{\mathbb{H}_{g}\ar[r]^{\mathfrak{M}_{G}}\ar[d]^{\mathbf{F}}&\mathbb{H}_{g}\ar[d]^{\mathbf{F}}
\\
{\rm Sym}_{>0}(2g,\mathbb{R})\ar[r]^{\mathfrak{C}_{G^{-1}}}&{\rm Sym}_{>0}(2g,\mathbb{R})}
\end{gathered}
\qquad
G\in\mathbf{Sp}(2g,\mathbb{Z}).
\end{gather*}
\end{itemize}
\end{Theorem}
For a~proof see~\cite[p.~148, Theorem~1]{SiegelIII}.

Now, since the original problem is concerned with the proper subset
$\mathcal{W}_{2g_0,1}\subset\mathbb{H}_{2g_0}$, let us introduce the space
\begin{gather*}
\mathcal{S}_{2g_0,1}:=\mathbf{F}(\mathcal{W}_{2g_0,1}).
\end{gather*}
In view of point ii) of Theorem~\ref{teo_Siegel}, it is easy to realize that $\mathfrak{C}$ can
be restricted to an action of the group $\mathbb{G}_{2g_0,1}$ on $\mathcal{S}_{2g_0,1}$.
Moreover the following diagram is commutative, for every $G$ belonging to $\mathbb{G}_{2g_0,1}$:
\begin{gather}\label{diag_comm_intro_1}
\begin{gathered}
\xymatrix{\mathcal{W}_{2g_0,1}\ar[r]^{\mathfrak{M}_{G}}\ar[d]^{\mathbf{F}}&\mathcal{W}_{2g_0,1}
\ar[d]^{\mathbf{F}}
\\
\mathcal{S}_{2g_0,1}\ar[r]^{\mathfrak{C}_{G^{-1}}}&\mathcal{S}_{2g_0,1}}
\end{gathered}
\qquad
G\in\mathbb{G}_{2g_0,1}.
\end{gather}
This immediately gives the following
\begin{Corollary}
\label{coro:Reform_1}
Let $\mathcal{D}^{\prime}$ be a~fundamental domain for the congruence action of the group
$\mathbb{G}_{2g_0,1}\subset\mathbf{Sp}(4g_0,\mathbb{Z})$ on $\mathcal{S}_{2g_0,1}$.
Then the set
$
\mathcal{D}:=\mathbf{F}^{-1}\left(\mathcal{D}^{\prime}\right)
$
is a~fundamental domain for the modular action of the same group $\mathbb{G}_{2g_0,1}$ on
$\mathcal{W}_{2g_0,1}$.
\end{Corollary}

In view of this, the original problem is equivalent to the quest of such a~$\mathcal{D}^{\prime}$.

To this purpose, a~more explicit characterization of $\mathcal{S}_{2g_0,1}$ will be useful:
\begin{Proposition}
The set $\mathcal{S}_{2g_0,1}$ consists of all real, symplectic, symmetric and positive definite matrices
$\Sigma$ of dimension $4g_0$, which have the following form
\begin{gather}\label{forma_Sigma}
\left[
\begin{matrix}\alpha&\beta&\gamma&\delta
\\
\beta&\alpha&-\delta&-\gamma
\\
\gamma^T&-\delta^T&\xi&\eta
\\
\delta^T&-\gamma^T&\eta&\xi
\end{matrix}
\right]
\in
\mathbf{Sp}(4g_0,\mathbb{R})\cap{\rm Sym}_{>0}(4g_0,\mathbb{R})
\end{gather}
for some $\alpha$, $\beta$, $\gamma$, $\delta$, $\xi$ and $\eta$ belonging to ${\rm Mat}(g_0,\mathbb{R})$.
\end{Proposition}

\begin{proof}
In view of Theorem~\ref{teo_Siegel}, point i), it will be suf\/f\/icient to prove that a~matrix
$\mathbf{w}$ of the Siegel upper half plane belongs to $\mathcal{W}_{2g_0,1}$ if and only if its image
$\mathbf{F}(\mathbf{w})$ has the form~\eqref{forma_Sigma}.
This last condition is equivalent to the relation
\begin{gather}
\label{equa:Riform_2}
T [\mathbf{F}(\mathbf{w}) ]T=\mathbf{F}(\mathbf{w}),
\end{gather}
where
\begin{gather}\label{equa:Riform_4}
T=\left[\begin{matrix}V&0\\0&-V\end{matrix}\right],
\qquad
V=\left[\begin{matrix}0&{\rm Id}\\{\rm Id}&0\end{matrix}\right].
\end{gather}
Using the explicit def\/inition~\eqref{def_Sigma}, one can show that~\eqref{equa:Riform_2} holds if and
only if the following relations do:
\begin{gather*}
V\lambda V=-\lambda,
\qquad
V\mu V=\mu.
\end{gather*}
But this happens if and only if $\mathbf{w}$ belongs to $\mathcal{W}_{2g_0,1}$, so the proof is complete.
\end{proof}

Next sections will be dedicated to determine a~fundamental domain $\mathcal{D}^{\prime}$ for the congruence
action of $\mathbb{G}_{2g_0,1}$ on $\mathcal{S}_{2g_0,1}$.

\section{Reduction toolkit}
\label{sez:ReductionToolkit}

In this section we introduce some objects which will be, so to say, useful tools in our task.

Let us start with an auxiliary map.
Consider the vector space\footnote{A more abstract, equivalent def\/inition of $\mathcal{V}_{2g_0,1}$ is the following:
\begin{gather*}
\mathcal{V}_{2g_0,1}:=\left\{\Sigma\in {\rm Mat}(4g_0,\mathbb{R})
\;
\text{such that}
\;
T\Sigma T=\Sigma\right\}.
\end{gather*}
Here $T$ denotes the idempotent matrix def\/ined in~\eqref{equa:Riform_4}.
}
\begin{gather*}
\mathcal{V}_{2g_0,1}:=\left\{\left[
\begin{matrix}
\alpha&\beta&\gamma&\delta
\\
\beta&\alpha&-\delta&-\gamma
\\
\pi&\rho&\xi&\eta
\\
-\rho&-\pi&\eta&\xi
\end{matrix}
\right]\in {\rm Mat}(4g_0,\mathbb{R})\right\}.
\end{gather*}
It is understood that the matrices $\alpha$, $\beta$, $\gamma$, $\delta$, $\pi$, $\rho$, $\xi$ and $\eta$ belong to
${\rm Mat}(g_0,\mathbb{R})$.
By a~straightforward calculation one verif\/ies that $\mathcal{V}_{2g_0,1}$ is closed w.r.t.\
matrix multiplication.

We def\/ine the map
\begin{gather*}
\mathbf{G}: \ \mathcal{V}_{2g_0,1}\longrightarrow {\rm Mat}(2g_0,\mathbb{R})\times {\rm Mat}(2g_0,\mathbb{R})
\end{gather*}
as follows:
\begin{gather}\label{equa:RedTool_01}
\mathbf{G}(\Sigma)=\left(\left[
\begin{matrix}
\alpha+\beta&\gamma-\delta
\\
\pi+\rho&\xi-\eta
\end{matrix}
\right],\left[
\begin{matrix}\alpha-\beta&\gamma+\delta
\\
\pi-\rho&\xi+\eta
\end{matrix}
\right]\right)
\qquad\forall\, \Sigma\in\mathcal{V}_{2g_0,1}.
\end{gather}
\begin{Lemma}
\label{lemma:RedTool_1}
Let $\Sigma$ and $\Sigma^{\prime}$ belong to $\mathcal{V}_{2g_0,1}$ and put
\begin{gather*}
\mathbf{G}(\Sigma)=(\sigma,\tau),
\qquad
\mathbf{G}\left(\Sigma^{\prime}\right)=\left(\sigma^{\prime},\tau^{\prime}\right).
\end{gather*}
One has the following:
\begin{itemize}\itemsep=0pt
\item[$i)$] $\mathbf{G}$ is an isomorphism of vector spaces from $\mathcal{V}_{2g_0,1}$ to
${\rm Mat}(2g_0,\mathbb{R})\times {\rm Mat}(2g_0,\mathbb{R})$.
\item[$ii)$] $\mathbf{G}\left( \Sigma\Sigma^{\prime}
\right)=\left(\sigma\sigma^{\prime},\tau\tau^{\prime}\right)$.
\item[$iii)$] $\mathbf{G}\left(\Sigma^{T}\right)=\left(\sigma^{T},\tau^{T}\right)$.
\item[$iv)$] $\Sigma$ is symmetric and
positive definite if and only if both $\sigma$ and $\tau$ are.
\item[$v)$] $\Sigma$ is symplectic if and only if
$\left(-J\tau^{T}J\right)\sigma={\rm Id}$.
\end{itemize}
\end{Lemma}

The matrix $J$ was def\/ined in~\eqref{definizione_simplettica_standard}.
\begin{proof}
Points i), ii) and iii) can be verif\/ied by means of a~straightforward
calculation.

iv) Since $\mathbf{G}$ is bijective (see point i)) one has that
\begin{gather*}
\Sigma=\Sigma^{T}\Longleftrightarrow\mathbf{G}(\Sigma)=\mathbf{G}\left(\Sigma^{T}\right)
\end{gather*}
but by point iii)
$
\mathbf{G}\left(\Sigma^{T}\right)=\left(\sigma^{T},\tau^{T}\right)$,
so
\begin{gather*}
\Sigma=\Sigma^{T}\Longleftrightarrow(\sigma,\tau)=\left(\sigma^{T},\tau^{T}\right)
\end{gather*}
and $\Sigma$ is symmetric if and only if both $\sigma$ and $\tau$ are; let us suppose that this is the case.
One can verify that the two identities
\begin{gather*}
\left(\begin{matrix}\mathbf{a}^{T}&\mathbf{b}^{T}\end{matrix}\right)
\sigma\left(\begin{matrix}\mathbf{a}\\\mathbf{b}\end{matrix}\right)
=\frac{1}{2}\left(\begin{matrix}\mathbf{a}^{T}&\mathbf{a}^{T}&\mathbf{b}^{T}&-\mathbf{b}^{T}\end{matrix}\right)
\Sigma\left(\begin{matrix}\mathbf{a}\\\mathbf{a}\\\mathbf{b}\\-\mathbf{b}\end{matrix}\right),
\\
\left(\begin{matrix}\mathbf{a}^{T}&\mathbf{b}^{T}\end{matrix}\right)
\tau\left(\begin{matrix}\mathbf{a}\\\mathbf{b}\end{matrix}\right)
=\frac{1}{2}\left(\begin{matrix}\mathbf{a}^{T}&-\mathbf{a}^{T}&\mathbf{b}^{T}&\mathbf{b}^{T}\end{matrix}\right)
\Sigma\left(\begin{matrix}\mathbf{a}\\-\mathbf{a}\\\mathbf{b}\\\mathbf{b}\end{matrix}\right)
\end{gather*}
hold true for every column vector $\mathbf{a}$, $\mathbf{b}$ belonging to $\mathbb{R}^{g_0}$.
This implies that if $\Sigma$ is positive def\/inite, then also $\sigma$ and $\tau$ are.

Viceversa, the quadratic form given by $\Sigma$ can be expressed in terms of the quadratic forms
corresponding to $\sigma$ and $\tau$ in the following way:
\begin{gather*}
\left(
\begin{matrix}\mathbf{a}^{T}&\mathbf{b}^{T}&\mathbf{c}^{T}&\mathbf{d}^{T}\end{matrix}
\right)
\Sigma\left(\begin{matrix}\mathbf{a}\\\mathbf{b}\\\mathbf{c}\\\mathbf{d}\end{matrix}\right)
=\frac{1}{2}
\left(\begin{matrix} (\mathbf{a}+\mathbf{b} )^{T}& (\mathbf{c}-\mathbf{d} )^{T}\end{matrix}\right)
\sigma\left(\begin{matrix} (\mathbf{a}+\mathbf{b} )\\ (\mathbf{c}-\mathbf{d} )\end{matrix}\right)
\\
\hphantom{\left(
\begin{matrix}\mathbf{a}^{T}&\mathbf{b}^{T}&\mathbf{c}^{T}&\mathbf{d}^{T}\end{matrix}
\right)
\Sigma\left(\begin{matrix}\mathbf{a}\\\mathbf{b}\\\mathbf{c}\\\mathbf{d}\end{matrix}\right)=}{}
+\frac{1}{2}
\left(\begin{matrix} (\mathbf{a}-\mathbf{b} )^{T}& (\mathbf{c}+\mathbf{d} )^{T}\end{matrix}\right)
\tau\left(\begin{matrix} (\mathbf{a}-\mathbf{b} )\\ (\mathbf{c}+\mathbf{d} )\end{matrix}\right).
\end{gather*}
So, if both $\sigma$ and $\tau$ are positive def\/inite, then also $\Sigma$ is.

 v)  By def\/inition, $\Sigma$ is symplectic if and only if
$
\Sigma^{T}J\Sigma=J$,
which, by carrying out the products in the left-hand side, reduces to the following system of
matrix-equations:
\begin{gather}
-\pi^{T}\alpha+\rho^{T}\beta+\alpha^{T}\pi - \beta^{T}\rho = 0,
\label{v:1}
\\
-\pi^{T}\beta+\rho^{T}\alpha+\alpha^{T}\rho - \beta^{T}\pi = 0,
\label{v:2}
\\
-\xi^{T}\gamma+\eta^{T}\delta +\gamma^{T}\xi -\delta^{T}\eta = 0,
\label{v:3}
\\
-\xi^{T}\delta+\eta^{T}\gamma+\gamma^{T}\eta -\delta^{T}\xi = 0,
\label{v:4}
\\
-\pi^{T}\gamma - \rho^{T}\delta+\alpha^{T}\xi+\beta^{T}\eta = {\rm Id},
\label{v:5}
\\
-\pi^{T}\delta -\rho^{T}\gamma+\alpha^{T}\eta +\beta^{T}\xi = {\rm Id}.
\label{v:6}
\end{gather}
Summing and subtracting (\ref{v:1}) and (\ref{v:2}) one gets the equations
\begin{gather}
- (\pi-\rho )^{T} (\alpha+\beta )+ (\alpha-\beta )^{T} (\pi+\rho )=0,
\nonumber
\\
- (\pi+\rho )^{T} (\alpha-\beta )+ (\alpha+\beta )^{T} (\pi-\rho )=0
\nonumber
\end{gather}
which are equivalent one to the other.
So ($\ref{v:1}$) and ($\ref{v:2}$) together are equivalent to the unique equation
\begin{gather*}
-(\pi-\rho)^{T}(\alpha+\beta)+(\alpha-\beta)^{T}(\pi+\rho)=0.
\end{gather*}
Analogously, ($\ref{v:3}$) and ($\ref{v:4}$) together are equivalent to
\begin{gather*}
-(\xi-\eta)^{T}(\gamma+\delta)+(\gamma-\delta)^{T}(\xi+\eta)=0.
\end{gather*}
Equations ($\ref{v:5}$) and ($\ref{v:6}$) are indeed equivalent to the system
\begin{gather*}
-(\pi+\rho)^{T}(\gamma+\delta)+(\alpha+\beta)^{T}(\xi+\eta)={\rm Id},
\\
-(\pi-\rho)^{T}(\gamma-\delta)+(\alpha-\beta)^{T}(\xi-\eta)={\rm Id}.
\end{gather*}
So $\Sigma$ is symplectic if and only if it satisf\/ies the following system of equations:
\begin{gather}
-(\pi-\rho)^{T}(\alpha+\beta)+(\alpha-\beta)^{T}(\pi+\rho)=0,
\nonumber
\\
-(\xi-\eta)^{T}(\gamma+\delta)+(\gamma-\delta)^{T}(\xi+\eta)=0,
\label{v:7}
\\
-(\pi+\rho)^{T}(\gamma+\delta)+(\alpha+\beta)^{T}(\xi+\eta)={\rm Id},
\label{v:8}
\\
-(\pi-\rho)^{T}(\gamma-\delta)+(\alpha-\beta)^{T}(\xi-\eta)={\rm Id},
\nonumber
\end{gather}
but after transposition of (\ref{v:7}) and (\ref{v:8}), this turns out to be equivalent to the condition
\begin{gather*}
\left(
\begin{matrix}
(\xi+\eta)^{T}&-(\gamma+\delta)^{T}
\\
-(\pi-\rho)^{T}&(\alpha-\beta)^{T}
\end{matrix}
\right)\left(
\begin{matrix}
(\alpha+\beta)&(\gamma-\delta)
\\
(\pi+\rho)&(\xi-\eta)
\end{matrix}
\right)={\rm Id},
\end{gather*}
which is simply
\begin{gather*}
\left(-J\tau^{T}J\right)\sigma={\rm Id}.\tag*{\qed}
\end{gather*}
\renewcommand{\qed}{}
\end{proof}

Let us now restrict to the group
\begin{gather*}
\mathbb{G}_{2g_0,1}(\mathbb{R})=\left\{\left[
\begin{matrix}
\alpha&\beta&\gamma&\delta
\\
\beta&\alpha&-\delta&-\gamma
\\
\pi&\rho&\xi&\eta
\\
-\rho&-\pi&\eta&\xi
\end{matrix}
\right]\in\mathbf{Sp}(4g_0,\mathbb{R})\right\}.
\end{gather*}
It is understood that the matrices $\alpha$, $\beta$, $\gamma$, $\delta$, $\pi$, $\rho$, $\xi$ and $\eta$ belong to
${\rm Mat}(g_0,\mathbb{R})$.
Notice that both $\mathbb{G}_{2g_0,1}$ and $\mathcal{S}_{2g_0,1}$ are contained in it.

Let us consider
\begin{gather*}
\mathbf{P}: \ \mathbb{G}_{2g_0,1}(\mathbb{R})\longrightarrow {\rm Mat}(2g_0,\mathbb{R}),
\qquad
\mathbf{P}(\Sigma)=\left[     
\begin{matrix}
\alpha+\beta&\gamma-\delta
\\
\pi+\rho&\xi-\eta
\end{matrix}
\right]
\qquad\forall\, \Sigma\in\mathbb{G}_{2g_0,1}(\mathbb{R}).
\end{gather*}
Many properties of this map can be deduced from Lemma~\ref{lemma:RedTool_1}:
\begin{Corollary}
\label{coro:RedTool_1}
The map $\mathbf{P}$ is a~homeomorphism from $\mathbb{G}_{2g_0,1}(\mathbb{R})$ to
$\mathbf{GL}(2g_0,\mathbb{R})$, this last one consisting of all real invertible matrices of
dimension $2g_0$.

Moreover it respects both matrix product
\begin{gather*}
\mathbf{P}\left(\Sigma\Sigma^{\prime}\right)=\mathbf{P}(\Sigma)\mathbf{P}\left(\Sigma^{\prime}
\right)
\qquad\forall\, \Sigma,\Sigma^{\prime}\in\mathbb{G}_{2g_0,1}(\mathbb{R})
\end{gather*}
and transposition
\begin{gather*}
\mathbf{P}\left(\Sigma^{T}\right)=\left[\mathbf{P}(\Sigma)\right]^{T}
\qquad\forall\, \Sigma\in\mathbb{G}_{2g_0,1}(\mathbb{R}).
\end{gather*}
\end{Corollary}

\begin{proof}
The inverse of $\mathbf{P}$ is the map
\begin{gather*}
\mathbf{Q}: \ \mathbf{GL}(2g_0,\mathbb{R})\longrightarrow\mathcal{S}_{2g_0,1}
\end{gather*}
def\/ined as follows:
\begin{gather*}
\mathbf{Q}(\sigma)=\mathbf{G}^{-1}\big(\sigma,-J\left(\sigma^{-1}\right)^{T}J\big)
\qquad\forall\, \sigma\in\mathbf{GL}(2g_0,\mathbb{R}).
\end{gather*}
Notice that the image of $\mathbf{Q}$ is contained in $\mathbb{G}_{2g_0,1}(\mathbb{R})$ because
of point~v) of the lemma.

Since both $\mathbf{Q}$ and $\mathbf{P}$ are continuous, this last one is a~homeomorphism.
Moreover $\mathbf{P}$ respects both matrix product and transposition due to point ii) and iii) of the lemma.
\end{proof}
\begin{Corollary}
\label{coro:RedTool_2}
The map $\mathbf{P}$ restricts to a~homeomorphism from $\mathcal{S}_{2g_0,1}$ to ${\rm
Sym}_{>0}(2g_0,\mathbb{R})$.
\end{Corollary}
\begin{proof}
From point iv) of Lemma~\ref{lemma:RedTool_1}, one immediately has that
\begin{gather*}
\mathbf{P}(\mathcal{S}_{2g_0,1})\subset{\rm Sym}_{>0}(2g_0,\mathbb{R}).
\end{gather*}
Viceversa, let $\sigma$ belong to ${\rm Sym}_{>0}(2g_0,\mathbb{R})$.
The matrix
\begin{gather*}
-J\left(\sigma^{-1}\right)^{T}J=J^{T}\left(\sigma^{-1}\right)^{T}J
\end{gather*}
is also positive def\/inite.

From the proof of the previous corollary, we know that
\begin{gather*}
\mathbf{P}^{-1}(\sigma)
=\mathbf{G}^{-1}\big(\sigma,-J\left(\sigma^{-1}\right)^{T}J\big)\in\mathbb{G}_{2g_0,1}(\mathbb{R}).
\end{gather*}
In view of point iv) of the lemma, this last matrix is also positive def\/inite and so it belongs
to~$\mathcal{S}_{2g_0,1}$.
It follows that
\begin{gather*}
\mathbf{P}(\mathcal{S}_{2g_0,1})={\rm Sym}_{>0}(2g_0,\mathbb{R})
\end{gather*}
and the proof is complete.
\end{proof}

Finally let us introduce the group
\begin{gather*}
\mathbb{K}_{2g_0,1}=\big\{g\in\mathbf{GL}(2g_0,\mathbb{Z})
\;
\text{such that}
\;
g^{T}J g\equiv J \ {\rm mod}\; 2\big\}.
\end{gather*}
One has the following
\begin{Corollary}
\label{coro:RedTool_3}
The map $\mathbf{P}$ restricts to an isomorphism of groups from $\mathbb{G}_{2g_0,1}$ to
$\mathbb{K}_{2g_0,1}$.
\end{Corollary}
\begin{proof}
Let $X$ belong to $\mathbb{G}_{2g_0,1}$ and put
$
\mathbf{G}(X)=(\sigma,\tau)$.
From point v) of the lemma, one has that
\begin{gather}\label{equa:RedTool_10}
\left(-J\tau^{T}J\right)\sigma={\rm Id}.
\end{gather}
Since both $\sigma$ and $\tau$ have integer coef\/f\/icients, this implies that
$
\sigma\in\mathbf{GL}(2g_0,\mathbb{Z})$.
Equation~\eqref{equa:RedTool_10} can be rewritten as follows:
\begin{gather}\label{equa:RedTool_20}
\tau^{T}J\sigma=J.
\end{gather}
On the other side, from def\/inition~\eqref{equa:RedTool_01} one has that
\begin{gather}\label{equa:RedTool_30}
\tau=\sigma-2\tilde{M}
\end{gather}
for some $\tilde{M}$ belonging to ${\rm Mat}(2g_0,\mathbb{Z})$, in this case.
Plugging~\eqref{equa:RedTool_30} into~\eqref{equa:RedTool_20} one obtains that
\begin{gather*}
\sigma^{T}J\sigma\equiv J
\mod 2.
\end{gather*}
Viceversa, let $g$ belong to $\mathbb{K}_{2g_0,1}$.
Let us put
\begin{gather*}
\mathbf{P}^{-1}(g)=\left[
\begin{matrix}
\alpha&\beta&\gamma&\delta
\\
\beta&\alpha&-\delta&-\gamma
\\
\pi&\rho&\xi&\eta
\\
-\rho&-\pi&\eta&\xi
\end{matrix}
\right],
\qquad
g\in\mathbb{K}_{2g_0,1}.
\end{gather*}
We need to prove that such inverse image has integer entries.
From the proof of Corollary~\ref{coro:RedTool_1} we know that
\begin{gather*}
\mathbf{P}^{-1} (g )=\mathbf{G}^{-1}\big(g,-J\left(g^{-1}\right)^{T}J\big),
\qquad
g\in\mathbb{K}_{2g_0,1}.
\end{gather*}
Recalling the def\/inition of $\mathbf{G}$ one can easily write
\begin{gather}\label{equa:RedTool_50}
\left[
\begin{matrix}\alpha&\gamma
\\
\pi&\xi
\end{matrix}
\right]=\frac{1}{2}\big[g-J\left(g^{-1}\right)^{T}J\big],
\qquad
\left[
\begin{matrix}\beta&-\delta
\\
\rho&-\eta
\end{matrix}
\right]=\frac{1}{2}\big[g+J\left(g^{-1}\right)^{T}J\big].
\end{gather}
Now, since
\begin{gather*}
g^{T}J g\equiv J
\mod 2
\end{gather*}
one also has that
\begin{gather}\label{equa:RedTool_60}
g+J\left(g^{-1}\right)^{T}J\equiv0
\mod 2.
\end{gather}
Since $g$ has an inverse with integer entries, it follows immediately that
\begin{gather}\label{equa:RedTool_70}
g-J\left(g^{-1}\right)^{T}J\equiv0
\mod 2.
\end{gather}
Equations~\eqref{equa:RedTool_50} together with~\eqref{equa:RedTool_60} and~\eqref{equa:RedTool_70} imply
that $\mathbf{P}^{-1}(g)$ has integer coef\/f\/icients.
So it belongs to $\mathbb{G}_{2g_0,1}$.
\end{proof}
\begin{Remark}
\label{rem:RedTool_1}
The whole modular group $\mathbf{Sp}(2g_0,\mathbb{Z})$ is contained in $\mathbb{K}_{2g_0,1}$.
For every matrix
\begin{gather*}
\left[\begin{matrix}A&C\\E&G\end{matrix}\right]\in\mathbf{Sp}(2g_0,\mathbb{Z})
\end{gather*}
one has that
\begin{gather*}
\mathbf{P}\left(\left[
\begin{matrix}
A&0&C&0
\\
0&A&0&-C
\\
E&0&G&0
\\
0&-E&0&G
\end{matrix}
\right]\right)=\left[
\begin{matrix}A&C
\\
E&G
\end{matrix}
\right].
\end{gather*}
It is easy to show that the argument of the function $\mathbf{P}$ in the left-hand side belongs to
$\mathbb{G}_{2g_0,1}$.

Since
\begin{gather*}
\left[\begin{matrix}{\rm Id}&0\\0&-{\rm Id}\end{matrix}\right]
\in
\mathbb{K}_{2g_0,1}
\end{gather*}
the following inclusion also holds:
\begin{gather*}
\mathbf{Sp}(2g_0,\mathbb{Z})\cdot\left[\begin{matrix}{\rm Id}&0\\0&-{\rm Id}\end{matrix}\right]
\subset
\mathbb{K}_{2g_0,1}.
\end{gather*}
For every
\begin{gather*}
\left[\begin{matrix}B&-D\\F&-H\end{matrix}\right]
\in
\mathbf{Sp}(2g_0,\mathbb{Z})\cdot
\left[\begin{matrix}{\rm Id}&0\\0&-{\rm Id}\end{matrix}\right]
\end{gather*}
one has that
\begin{gather*}
\mathbf{P}\left(\left[\begin{matrix}0&B&0&D\\B&0&-D&0\\0&F&0&H\\-F&0&H&0\end{matrix}\right]\right)
=\left[\begin{matrix}B&-D\\F&-H\end{matrix}\right]
\end{gather*}
and the argument of $\mathbf{P}$ again belongs to $\mathbb{G}_{2g_0,1}$.
\end{Remark}

\section{Reduction of the problem}
\label{sez:Reduction}

We can now reformulate further the original problem, using the instruments developed in the previous
section.
This process bears the name of reduction because it halves the dimension of the matrices involved in the
issue.
\begin{Proposition}
\label{prop:RedProb_1}
The following diagram
\begin{gather}\label{diag_comm_CR}
\begin{gathered}
\xymatrix{
\mathcal{S}_{2g_0,1}\ar[r]^{\mathfrak{C}_{G}}\ar[d]^{\mathbf{P}}
& \mathcal{S}_{2g_0,1}\ar[d]^{\mathbf{P}}
\\
{\rm Sym}_{>0}(2g_0,\mathbb{R})\ar[r]^{\mathfrak{C}_{\mathbf{P}(G)}}
& {\rm Sym}_{>0}(2g_0,\mathbb{R})
}
\end{gathered}
\qquad
G\in\mathbb{G}_{2g_0,1}
\end{gather}
is commutative for every $G$ belonging to $\mathbb{G}_{2g_0,1}$.
\end{Proposition}
\begin{proof}
Let $\Sigma$ and $G$ belong to $\mathcal{S}_{2g_0,1}$ and $\mathbb{G}_{2g_0,1}$ respectively.
Due to Corollary~\ref{coro:RedTool_1} one has
\begin{gather*}
\mathbf{P} [\mathfrak{C}(\Sigma,G) ]=\mathbf{P}\left[G^{T}
\Sigma G\right]= [\mathbf{P}(G) ]^{T}\mathbf{P}(\Sigma)\mathbf{P}(G),
\qquad
\Sigma\in\mathcal{S}_{2g_0,1},
\qquad
G\in\mathbb{G}_{2g_0,1}.
\end{gather*}
On the other side, simply by def\/inition,
\begin{gather*}
\mathfrak{C} [\mathbf{P}(\Sigma),\mathbf{P}(G) ]
= [\mathbf{P}(G) ]^{T}\mathbf{P}(\Sigma)\mathbf{P}(G),
\qquad
\Sigma\in\mathcal{S}_{2g_0,1},
\qquad
G\in\mathbb{G}_{2g_0,1}.\tag*{\qed}
\end{gather*}
\renewcommand{\qed}{}
\end{proof}

As a~consequence, one has the following
\begin{Theorem}
Let $\mathcal{D}^{\prime\prime}$ be a~fundamental domain for the congruence action of $\mathbb{K}_{2g_0,1}$
on\linebreak
${\rm Sym}_{>0}(2g_0,\mathbb{R})$.
Then the set
$
\mathcal{D}^{\prime}:=\mathbf{P}^{-1}\left(\mathcal{D}^{\prime\prime}\right)
$
is a~fundamental domain for the congruence action of $\mathbb{G}_{2g_0,1}$ on $\mathcal{S}_{2g_0,1}$.
\end{Theorem}

\begin{proof}
The proof is a~standard argument based on Proposition~\ref{prop:RedProb_1} and
Corollaries~\ref{coro:RedTool_2} \linebreak and~\ref{coro:RedTool_3}.
\end{proof}

The original problem is thus equivalent to the quest of such a~$\mathcal{D}^{\prime\prime}$.
The remaining sections will be dedicated to this issue.
This will also emphasize the full advantage of our approach.

Another consequence of Proposition~\ref{prop:RedProb_1} is worth pointing out.
For every $\mathbf{w}$ belonging to~$\mathcal{W}_{2g_0,1}$, let us def\/ine
\begin{gather}
\label{equa:RedProb_100}
\mathcal{I}(\mathbf{w})
=\det\big[ (\mathbf{P}\circ\mathbf{F} )(\mathbf{w})\big],
\qquad
\mathbf{w}\in\mathcal{W}_{2g_0,1}.
\end{gather}

\begin{Corollary}
\label{coro:RedProb_1}
Let $\mathbf{w}$ and $G$ belong to $\mathcal{W}_{2g_0,1}$ and $\mathbb{G}_{2g_0,1}$ respectively.
Consider the modular transformation
\begin{gather*}
\tilde{\mathbf{w}}=\mathfrak{M}(G,\mathbf{w}).
\end{gather*}
One has
\begin{gather*}
\mathcal{I}(\tilde{\mathbf{w}})=\mathcal{I}(\mathbf{w}).
\end{gather*}
\end{Corollary}

\begin{proof}
In view of commutative diagram~\eqref{diag_comm_intro_1},
\begin{gather*}
\mathbf{F} [\mathfrak{M}(G,\mathbf{w}) ]
=\mathfrak{C}\big[\mathbf{F}(\mathbf{w}),G^{-1}\big],
\qquad
\mathbf{w}\in\mathcal{W}_{2g_0,1},
\qquad
G\in\mathbb{G}_{2g_0,1}.
\end{gather*}
Taking the image via $\mathbf{P}$ of both sides, and applying Proposition~\ref{prop:RedProb_1}
\begin{gather*}
 (\mathbf{P}\circ\mathbf{F} ) [\mathfrak{M}(G,\mathbf{w}) ]
=\mathfrak{C}\big[ (\mathbf{P}\circ\mathbf{F} )(\mathbf{w}),\mathbf{P}\big(G^{-1}\big)\big].
\end{gather*}
Considering the determinants, and recalling the explicit def\/inition~\eqref{def_cong_2},
\begin{gather*}
\det \{ (\mathbf{P}\circ\mathbf{F} ) [\mathfrak{M}(G,\mathbf{w}) ] \}
=\det\big[\mathbf{P}\big(G^{-1}\big)\big]^{2}
\det [ (\mathbf{P}\circ\mathbf{F} )(\mathbf{w}) ].
\end{gather*}
The thesis follows from this last equation, in view of the fact that $\mathbb{K}_{2g_0,1}$ is contained in
$\mathbf{GL}(2g_0,\mathbb{Z})$.
\end{proof}

From Corollary~\ref{coro:RedTool_2}, one has
\begin{gather*}
 (\mathbf{P}\circ\mathbf{F} )(\mathcal{W}_{2g_0,1})={\rm Sym}_{>0}(2g_0,\mathbb{R}).
\end{gather*}
As a~consequence, the quantity $\mathcal{I}$ is not constant over the whole $\mathcal{W}_{2g_0,1}$.
In other words, it is not a~trivial invariant with respect to the modular action $\mathfrak{M}$ of the
group $\mathbb{G}_{2g_0,1}\subset \mathbf{Sp}(4g_0,\mathbb{Z})$.

To the best of our understanding, no geometrical interpretation of this quantity is available so far.

\section[The case $2g_0=2$]{The case $\boldsymbol{2g_0=2}$}
\label{sez:CaseGenusTwo}

We start here to put in concrete action the previously developed abstract theory.
The simplest case when $2g_0$ equals 2 allows an elegant solution, together with a~deeper understanding of
the structure of the problem.
\begin{Lemma}
One has
\begin{gather}
\label{equa:GenTwo_1}
\mathbb{K}_{2,1}=\mathbf{GL}(2,\mathbb{Z}).
\end{gather}
Moreover, if $G$ belongs to $\mathbb{G}_{2,1}$, there are only two possible cases: either
\begin{gather}\label{equa:GenTwo_2}
G=\left[\begin{matrix}a&0&c&0\\0&a&0&-c\\e&0&g&0\\0&-e&0&g\end{matrix}\right],
\qquad
ag-ec=1
\end{gather}
or
\begin{gather}\label{equa:GenTwo_3}
G=\left[\begin{matrix}0&b&0&d\\b&0&-d&0\\0&f&0&h\\-f&0&h&0\end{matrix}\right],
\qquad
bh-fd=1.
\end{gather}
\end{Lemma}
\begin{proof}
As pointed out in Remark~\ref{rem:RedTool_1}, one has the following inclusions:   
\begin{gather*}
\mathbf{Sp}(2,\mathbb{Z})\subset\mathbb{K}_{2,1},
\qquad
\mathbf{Sp}(2,\mathbb{Z})\cdot\left[\begin{matrix}1&0\\0&-1\end{matrix}\right]\subset\mathbb{K}_{2,1}.
\end{gather*}
On the other side,
\begin{gather*}
\mathbf{GL}(2,\mathbb{Z})
=\mathbf{Sp}(2,\mathbb{Z})\cup\mathbf{Sp}(2,\mathbb{Z})\cdot
\left[\begin{matrix}1&0\\0&-1\end{matrix}\right].
\end{gather*}
This gives~\eqref{equa:GenTwo_1}.

Now, let $G$ belong to $\mathbb{G}_{2,1}$.
Then, either
\begin{gather}
\label{equa:GenTwo_4}
\mathbf{P}(G)\in\mathbf{Sp}(2,\mathbb{Z})
\end{gather}
or
\begin{gather}\label{equa:GenTwo_5}
\mathbf{P}(G)\in\mathbf{Sp}(2,\mathbb{Z})\cdot\left[\begin{matrix}1&0\\0&-1\end{matrix}\right].
\end{gather}
Always in view of Remark~\ref{rem:RedTool_1},~\eqref{equa:GenTwo_4} and~\eqref{equa:GenTwo_5} correspond
to~\eqref{equa:GenTwo_2} and~\eqref{equa:GenTwo_3} respectively.
\end{proof}

By means of~\eqref{equa:GenTwo_1} we are reconducted to a~classical problem whose solution is well-known.
Let us indicate with
\begin{gather*}
\left[\begin{matrix}\phi&\chi\\\chi&\psi\end{matrix}\right]\in{\rm Sym}_{>0}(2,\mathbb{R})
\end{gather*}
the generic element of ${\rm Sym}_{>0}(2,\mathbb{R})$.
A fundamental domain $\mathcal{D}^{\prime\prime}$ for the congruence action of the group $\mathbb{K}_{2,1}
= \mathbf{GL}(2,\mathbb{Z})$ on this space is described by the following system of inequalities{\samepage
\begin{gather}\label{Lagrange}
\mathcal{D}^{\prime\prime}:
\
\phi\leq\psi,
\qquad
-\phi\leq2\chi\leq0.
\end{gather}
This result was already known to Lagrange and Hermite.}

Let us denote with
\begin{gather*}
\mathbf{w}=\left(
\begin{matrix}\gamma+\imath\delta&\imath\beta
\\
\imath\beta&-\gamma+\imath\delta
\end{matrix}
\right)\in\mathcal{W}_{2,1},
\qquad
\beta,\gamma,\delta\in\mathbb{R},
\end{gather*}
the generic element of $\mathcal{W}_{2,1}$.
One might recall that some such~$\mathbf{w}$'s are period matrices of a~genus two real Riemann surface.
It is possible to prove that if this last one is separated, with just one oval, then $\beta > 0$.
If instead there are no f\/ix points for~$r$, then $\beta < 0$.
Finally, no smooth curve can yield $\beta=0$.
(These inequalities can be directly deduced from Lemma~10.10 of~\cite{SilCom}, via an appropriate change of
basis in the homology.)

The imaginary part of $\mathbf{w}$ is positive def\/inite if and only if
$
\delta>|\beta|$.
The map $\mathbf{P}\circ\mathbf{F}$ has the following simple expression in this case:
\begin{gather}
\label{composta_esplicita}
 (\mathbf{P}\circ\mathbf{F} )(\mathbf{w})=\frac{1}{\delta+\beta}
\left(\begin{matrix}1&-\gamma\\-\gamma&\gamma^2+\delta^2-\beta^2\end{matrix}\right).
\end{gather}
These simple facts, together with the results of the previous sections, suf\/f\/ice to easily reobtain the
following result already known to Silhol~\cite{SilAlg}.
\begin{Theorem}
A fundamental domain $\mathcal{D}$ for the modular action $\mathfrak{M}$ of the group
$\mathbb{G}_{2,1}\subset \mathbf{Sp}(4,\mathbb{Z})$ over $\mathcal{W}_{2,1}$ is described by the
following system of inequalities:
\begin{gather}\label{Risultato_g2}
\mathcal{D}:
\
1\leq\gamma^2+\delta^2-\beta^2,
\qquad
0\leq\gamma\leq\frac{1}{2},
\qquad
\delta>|\beta|.
\end{gather}
\end{Theorem}
\begin{proof}
Due to Corollaries~\ref{coro:RedProb_1} and~\ref{coro:Reform_1}, such a~fundamental domain $\mathcal{D}$
can be found by simply considering
\begin{gather*}
\mathcal{D}= (\mathbf{P}\circ\mathbf{F} )^{-1}\left(\mathcal{D}^{\prime\prime}\right).
\end{gather*}
In view of~\eqref{composta_esplicita}, one gets an explicit description of this set operating the
substitution
\begin{gather*}
\phi=\frac{1}{\delta+\beta},
\qquad
\chi=-\frac{\gamma}{\delta+\beta},
\qquad
\psi=\frac{\gamma^2+\delta^2-\beta^2}{\delta+\beta}
\end{gather*}
into the system~\eqref{Lagrange}.
This leads to~\eqref{Risultato_g2}.
\end{proof}
With this result, the original problem can be considered completely solved when $2g_0$ equals $2$.

The theory developed up to now, though, allows a~deeper insight into the structure of the modular action of
$\mathbb{G}_{2,1}$ on $\mathcal{W}_{2,1}$.
\begin{Theorem}
\label{vento}
Let us introduce the following system of coordinates on the space $\mathcal{W}_{2,1}$:
\begin{gather}
\label{equa:GenTwo_10}
\mathcal{I}(\mathbf{w})=\frac{\delta-\beta}{\delta+\beta},
\qquad
\boldsymbol{\tau}(\mathbf{w})=\gamma+\imath\sqrt{\delta^2-\beta^2},
\qquad
\mathbf{w}\in\mathcal{W}_{2,1}.
\end{gather}
The square root is chosen to be positive.

Consider the modular transformation given by an element of $\mathbb{G}_{2,1}$:
\begin{gather*}
\tilde{\mathbf{w}}=\mathfrak{M}(G,\mathbf{w}),
\qquad
\mathbf{w}\in\mathcal{W}_{2,1},
\qquad
G\in\mathbb{G}_{2,1}.
\end{gather*}
In terms of the new coordinates, this acts as follows:
\begin{gather}
\label{gia_fatta}
\mathcal{I}(\tilde{\mathbf{w}})=\mathcal{I}(\mathbf{w})
\end{gather}
and
\begin{gather}
\label{equa:GenTwo_20}
\boldsymbol{\tau}(\tilde{\mathbf{w}})=\frac{a\boldsymbol{\tau}(\mathbf{w})+c}
{e\boldsymbol{\tau}(\mathbf{w})+g}
\end{gather}
if $G$ has the form~\eqref{equa:GenTwo_2}, while
\begin{gather}
\label{equa:GenTwo_30}
\boldsymbol{\tau}(\tilde{\mathbf{w}})=\frac{b\overline{\boldsymbol{\tau}(\mathbf{w})}
-d}{f\overline{\boldsymbol{\tau}(\mathbf{w})}-h}
\end{gather}
if~\eqref{equa:GenTwo_3} holds.
\end{Theorem}
\begin{proof}
The explicit expression for $\mathcal{I}(\mathbf{w})$ in~\eqref{equa:GenTwo_10} agrees with its
general def\/inition given in Section~\ref{sez:Reduction}.
To see this it is suf\/f\/icient to plug~\eqref{composta_esplicita} into~\eqref{equa:RedProb_100}.
Formula~\eqref{gia_fatta} is just the content of Corollary~\ref{coro:RedProb_1}.

Gluing together~\eqref{diag_comm_intro_1} and~\eqref{diag_comm_CR} one gets the following commutative
diagram:
\begin{gather}\label{incollato}
\begin{gathered}
\xymatrix{\mathcal{W}_{2g_0,1}\ar[r]^{\mathfrak{M}_{G}}\ar[d]_{\mathbf{\mathbf{P}\circ F}}&\mathcal{W}
_{2g_0,1}\ar[d]^{\mathbf{P}\circ\mathbf{F}}
\\
{\rm Sym}_{>0}(2g_0,\mathbb{R})\ar[r]^{\mathfrak{C}_{\mathbf{P}\left(G^{-1}\right)}}&{\rm Sym}
_{>0}(2g_0,\mathbb{R})}
\end{gathered}
\qquad
G\in\mathbb{G}_{2,1}.
\end{gather}

Let us introduce
\begin{gather*}
\mathbf{u}(\mathbf{w})
:=\frac{1}{\sqrt{\mathcal{I}(\mathbf{w})}}
 [ (\mathbf{P}\circ\mathbf{F} )(\mathbf{w}) ],
\qquad
\mathbf{w}\in\mathcal{W}_{2,1},
\end{gather*}
where the square root is again chosen to be positive.
Using~\eqref{incollato} one can write
\begin{gather}
\mathbf{u}(\tilde{\mathbf{w}})=\frac{1}{\sqrt{\mathcal{I}(\tilde{\mathbf{w}})}}
\left[\left(\mathbf{P}\circ\mathbf{F}\right)(\tilde{\mathbf{w}})\right]
=\frac{1}{\sqrt{\mathcal{I}(\tilde{\mathbf{w}})}}\left\{\mathfrak{C}\left[\left(\mathbf{P}
\circ\mathbf{F}\right)(\mathbf{w}),\mathbf{P}\left(G^{-1}\right)\right]\right\}
\nonumber
\\
\phantom{\mathbf{u}(\tilde{\mathbf{w}})}{}
=\left[\mathbf{P}\left(G^{-1}\right)\right]^{T}\left\{\frac{1}{\sqrt{\mathcal{I}(\mathbf{w})}}
\left[\left(\mathbf{P}\circ\mathbf{F}\right)(\mathbf{w})\right]\right\}\left[\mathbf{P}
\left(G^{-1}\right)\right]
\nonumber
\\
\phantom{\mathbf{u}(\tilde{\mathbf{w}})}{}
=\left[\mathbf{P}\left(G^{-1}\right)\right]^{T}\left[\mathbf{u}\left(\mathbf{w}
\right)\right]\left[\mathbf{P}\left(G^{-1}\right)\right].
\label{equa:GenTwo_110}
\end{gather}
In view of def\/inition~\eqref{equa:RedProb_100}, one has that
\begin{gather*}
\det(\mathbf{u})=1
\qquad\forall\, \mathbf{w}\in\mathcal{W}_{2,1}.
\end{gather*}
This means that $\mathbf{u}(\mathbf{w})$ is not only symmetric and positive def\/inite but also
symplectic for every~$\mathbf{w}$ belonging to~$\mathcal{W}_{2,1}$.
As a~consequence of Theorem~\ref{teo_Siegel}, then, there exists a~unique point
$\boldsymbol{\tau}(\mathbf{w})$ in the Siegel upper half plane of degree one such that
\begin{gather*}
\mathbf{F} (\boldsymbol{\tau}(\mathbf{w}) )=\mathbf{u}(\mathbf{w}),
\qquad
\mathbf{w}\in\mathcal{W}_{2,1}.
\end{gather*}
Using~\eqref{def_Sigma}, \eqref{composta_esplicita} and~\eqref{equa:RedProb_100} one can verify that this
def\/inition of $\boldsymbol{\tau}$ coincides with the more explicit one given in~\eqref{equa:GenTwo_10}.

Now, suppose that $G$ has the form~\eqref{equa:GenTwo_2}.
In this case,
\begin{gather*}
\mathbf{P}(G)=\left[\begin{matrix}a&c\\e&g\end{matrix}\right]\in\mathbf{Sp}(2,\mathbb{Z}).
\end{gather*}
By point ii) of Theorem~\ref{teo_Siegel},~\eqref{equa:GenTwo_110} gives~\eqref{equa:GenTwo_20}.

Let $G$ have the form~\eqref{equa:GenTwo_3}, instead.
One has that
\begin{gather*}
\mathbf{P}(G)\left[\begin{matrix}1&0\\0&-1\end{matrix}\right]
=\left[\begin{matrix}b&d\\f&h\end{matrix}\right]\in\mathbf{Sp}(2,\mathbb{Z}).
\end{gather*}
Equation~\eqref{equa:GenTwo_110} can be rewritten as follows:
\begin{gather}
\label{equa:GenTwo_40}
\mathbf{u}(\tilde{\mathbf{w}})
=\left[\left(\mathbf{P}(G)\left[\begin{matrix}1&0\\0&-1\end{matrix}\right]\right)^{-1}\right]^{T}
\left(\left[\begin{matrix}1&0\\0&-1\end{matrix}\right]
\mathbf{u}(\mathbf{w})\left[\begin{matrix}1&0\\0&-1\end{matrix}\right]\right)
\left(\mathbf{P}(G)\left[\begin{matrix}1&0\\0&-1\end{matrix}\right]\right)^{-1}.
\end{gather}
By means of~\eqref{def_Sigma}, one can verify that
\begin{gather*}
\mathbf{F}\left(-\overline{\boldsymbol{\tau}(\mathbf{w})}\right)
=\left[\begin{matrix}1&0\\0&-1\end{matrix}\right]\mathbf{u}(\mathbf{w})
\left[\begin{matrix}1&0\\0&-1\end{matrix}\right].
\end{gather*}
Again by point ii) of Theorem~\ref{teo_Siegel} equality~\eqref{equa:GenTwo_40}
implies~\eqref{equa:GenTwo_30}
\end{proof}

The previous theorem has some interesting geometrical consequences.
\begin{Observation}
Let $(\Gamma, r)$ be a~genus two, real Riemann surface of separated type with just one oval,
or a~genus two real Riemann surface with no invariant points for~$r$.
Let us fix a~basis in its homology of the form~\eqref{Intro_1}--\eqref{Intro_3}.
One can then compute a~period matrix $\mathbf{w}$ and the corresponding $\boldsymbol{\tau} =
\boldsymbol{\tau} ( \mathbf{w}  ) \in \mathbb{H}_{1}$.
There exists a~unique such $\boldsymbol{\tau}$, say $\boldsymbol{\tau}_{0}$, belonging to the set
\begin{gather*}
\mathcal{D}_{\boldsymbol{\tau}}:=\left\{\boldsymbol{\tau}\in\mathbb{H}_{1}
\;
\text{such that}
\;
0\leq\Re (\boldsymbol{\tau} )\leq\frac{1}{2},\,1\leq |\boldsymbol{\tau} |\right\}.
\end{gather*}
This fact can be verified using Theorem~{\rm \ref{vento}} together with the classical theory of M\"obius
transformations.
{}$\mathcal{D}_{\boldsymbol{\tau}}$ is a~fundamental domain for the action of $\mathbb{G}_{2,1} =
\mathbf{GL} ( 2,\mathbb{Z} )$ on~$\mathbb{H}_1$ defined
by~\eqref{equa:GenTwo_20}, \eqref{equa:GenTwo_30}.
Moreover, no two distinct points lying on the boundary of $\mathcal{D}_{\boldsymbol{\tau}}$ belong to the
same orbit.

To every real Riemann surface as above, then, there corresponds a~couple
\begin{gather}
\label{Zafira}
(\Gamma,r)\longrightarrow (\mathcal{I},\boldsymbol{\Delta} ),
\end{gather}
where $\mathcal{I}\in \mathbb{R}$ is the quantity introduced in~\eqref{equa:GenTwo_10} and
$\boldsymbol{\Delta}$ is the genus one Riemann surface whose period matrix is $\boldsymbol{\tau}_{0}$.

Let us remark that these objects do not depend on the particular basis in the homology of~$\Gamma$ used to
compute them.
To the best of our knowledge, correspondence~\eqref{Zafira} never appeared in the literature before.
A geometric characterization of~$\mathcal{I}$ and~$\boldsymbol{\Delta}$ independent of the period matrix~$\mathbf{w}$ is not available yet and it is meant to be part of a~work in progress.
\end{Observation}

\section{The general case}\label{sez:GeneralCase}

At the beginning of the last century Minkowski studied the congruence action of
$\mathbf{GL}(n,\mathbb{Z})$ on ${\rm Sym}_{>0}(n,\mathbb{R})$: he exhibited, for
every $n\geq 2$, a~fundamental domain $\mathcal{M}$ which can be described by a~f\/inite set of inequalities
\begin{gather*}
f_{1}(\sigma)\geq0,
\qquad
f_{2}(\sigma)\geq0,
\qquad
\ldots,
\qquad
f_{m_n}(\sigma)\geq0,
\qquad
\sigma\in{\rm Sym}_{>0}(n,\mathbb{R}),
\end{gather*}
where $f_{1},f_{2},\ldots,f_{m_n}$ are \textit{linear homogeneous expressions} of the entries of $\sigma$.
For low dimensions these expressions were explicitly determined: see~\cite{Barnes-Cohn} and~\cite{Ryshkov}.

After considerations of the previous sections, we now need a~fundamental domain for the congruence action,
on ${\rm Sym}_{>0}(2g_0,\mathbb{R})$, of the group $\mathbb{K}_{2g_0,1}$, which is strictly
contained in $\mathbf{GL}(2g_0,\mathbb{Z})$ for $2g_0\geq 4$.
The index of this subgroup is f\/inite.
In particular
\begin{Lemma}
One has
\begin{gather}
\label{indice_esplicito}
\left[\mathbf{GL}(2g_0,\mathbb{Z}):\mathbb{K}_{2g_0,1}\right]=2^{g_0\left(g_0-1\right)}
\prod_{j=1}^{g_0}\left(2^{2j-1}-1\right).
\end{gather}
\end{Lemma}
\begin{proof}
Consider the quotient map
\begin{gather*}
\phi
:
\
\mathbf{GL}(2g_0,\mathbb{Z})\longrightarrow\mathbf{GL}(2g_0,\mathbb{Z}_2),
\end{gather*}
which associates to every unimodular matrix its class modulo 2 elementwise: it is surjective and
\begin{gather*}
\phi(\mathbb{K}_{2g_0,1})=\mathbf{Sp}(2g_0,\mathbb{Z}_{2})
\end{gather*}
(see~\cite[Lemma~4]{Newman-Reiner}  and~\cite[Theorem~1]{Newman-Smart}).
Moreover
\begin{gather*}
\Ker(\phi)\subset\mathbb{K}_{2g_0,1}.
\end{gather*}
Because of these facts $\phi$ induces a~bijection between the left cosets of $\mathbb{K}_{2g_0,1}$ in
$\mathbf{GL}(2g_0,\mathbb{Z})$ and the ones of $\mathbf{Sp(2g_0,\mathbb{Z}_2)}$ in
$\mathbf{GL}(2g_0,\mathbb{Z}_{2})$.
So,
\begin{gather*}
\left[\mathbf{GL}(2g_0,\mathbb{Z}):\mathbb{K}_{2g_0,1}\right]
= {\frac{\card\left(\mathbf{GL}(2g_0,\mathbb{Z}_{2})\right)}
{\card\left(\mathbf{Sp}(2g_0,\mathbb{Z}_2)\right)}}.
\end{gather*}
Formula~\eqref{indice_esplicito} follows from this last equality keeping into account that
(see~\cite{ConAtl})
\begin{gather*}
\mathbf{GL}(2g_0,\mathbb{Z}_{2})=2^{g_0\left(2g_0-1\right)}\prod_{j=1}^{2g_0}\left(2^{j}-1\right)
\end{gather*}
and that
\begin{gather*}
\mathbf{Sp}(2g_0,\mathbb{Z}_{2})=2^{g_0^2}\prod_{j=1}^{g_0}\left(2^{2j}-1\right).\tag*{\qed}
\end{gather*}
\renewcommand{\qed}{}
\end{proof}

Our technique to determine a~fundamental domain for the congruence action of the group $\mathbb{K}_{2g_0,1}$ on~${\rm Sym}_{>0}( 2g_0, \mathbb{R})$ is based on the following
\begin{Proposition}\label{proposizione_prima_cinque}
Let us consider the left cosets of $\mathbb{K}_{2g_0,1}$ in $\mathbf{GL}(2g_0,\mathbb{Z})$
\begin{gather*}
L_1,\quad L_2,\quad \ldots,\quad L_m,
\end{gather*}
and fix a~representative for each of them:
\begin{gather*}
g_1\in L_1,\quad g_2\in L_2,\quad \ldots,\quad g_m\in L_m.
\end{gather*}
If
\begin{gather}\label{Unione}
\text{\rm int}\left(\bigcup_{i=1}^{m}\mathfrak{C}(\mathcal{M},g_i)\right)
\end{gather}
is a~connected set then
\begin{gather}
\label{insieme_unione}
\bigcup_{i=1}^{m}\mathfrak{C}(\mathcal{M},g_i)
\end{gather}
is a~fundamental domain for the congruence action of $\mathbb{K}_{2g_0,1}$ on ${\rm
Sym}_{>0}(2g_0,\mathbb{R})$.
\end{Proposition}

Similar results were used to compute fundamental domains in other contexts (see for example~\cite[Theorem 5.2]{LasOri}).
Since we are not aware of any appropriate reference for our specif\/ic case, we provide a~proof of it here below.
\begin{notation}
Let $g$ be an element of $\mathbf{GL}(n,\mathbb{Z})$.
Throughout this section, the symbol $\mathfrak{C}_{g}$ is understood to denote a~map from the whole ${\rm
Sym}_{>0}(n,\mathbb{R})$ into itself.
\end{notation}

\begin{Lemma}
\label{lem_primo_cinque}
Let $\sigma$ be an interior point of $\mathcal{M}$, and $g\in \mathbf{GL}(n,\mathbb{Z})$ such
that $\mathfrak{C}(\sigma,g)\in \mathcal{M}$.
Then $g = \pm {\rm Id}$.
\end{Lemma}
\begin{proof}
[Proof of Lemma~\ref{lem_primo_cinque}] By def\/inition of fundamental domain, $\mathfrak{C}_g$ needs to
keep $\sigma$ f\/ixed.
Due to continuity, then, there exists an open neighbourhood of $\sigma$ whose image is contained in
$\mathcal{M}$.
The map $\mathfrak{C}_g$ reduces to the identity on such neighborhood and so on the whole ${\rm
Sym}_{>0}(n,\mathbb{R})$, because it is linear.
This implies that $g = \pm {\rm Id}$.
\end{proof}

\begin{proof}[Proof of Proposition~\ref{proposizione_prima_cinque}] The nontrivial part of the proof is the
verif\/ication of point ii) of Def\/inition~\ref{def_dom_fond}.

Let $\rho$ be an interior point of $\bigcup_{i=1}^{m}\mathfrak{C}(\mathcal{M},g_i)$.
Suppose that
\begin{gather*}
\rho=\mathfrak{C}(\sigma,g)
\end{gather*}
for some $g$ belonging to $\mathbf{GL}(2g_0,\mathbb{Z})$ and some $\sigma$ in
$\mathfrak{C}(\mathcal{M},g_1)$.
By continuity of $\mathfrak{C}_g$, there exists an interior point $\tau$ of $\mathfrak{C} (\mathcal{M},g_1  )$ such that
\begin{gather}
\label{dim_compatta_uno}
\pi=\mathfrak{C} (\tau,g )
\end{gather}
is also an interior point of~\eqref{insieme_unione}.
There exist $\tau_0$ and $\pi_0$ belonging to $\mathcal{M}$ such that
\begin{gather}
\label{dim_compatta_due}
\pi=\mathfrak{C} (\pi_0,g_k ),
\qquad
\tau=\mathfrak{C} (\tau_0,g_1 )
\end{gather}
for some $1\leq k\leq m$.
Formulas~\eqref{dim_compatta_uno} and~\eqref{dim_compatta_due} together give
\begin{gather*}
\pi_0=\mathfrak{C}\big(\tau_0,g_1g g_k^{-1}\big).
\end{gather*}
In view of Lemma~\ref{lem_primo_cinque} this yields
\begin{gather*}
g_1g g_k^{-1}=\pm{\rm Id}
\
\Longrightarrow
\
g=\pm g_{1}^{-1}g_k.
\end{gather*}
Now, if $k$ equals one, then $g = \pm {\rm Id}$ and $\sigma$ coincides with $\rho$.
Otherwise~$g$ cannot belong to~$\mathbb{K}_{2g_0,1}$, because $g_1$ and $g_k$ come from distinct cosets by
construction.
The cases when~$\sigma$ belongs to $\mathfrak{C} ( \mathcal{M},g_j  )$, with $2\leq j\leq m$ can
be treated analogously.
One concludes that an interior point~$\rho$ cannot be equivalent (in the $\mathbb{K}_{2g_0,1}$-sense) to
any other point of $\bigcup_{i=1}^{m}\mathfrak{C} (\mathcal{M},g_i )$.
\end{proof}

After Proposition~\ref{proposizione_prima_cinque}, the issue reduces to select a~set of representatives
such that the set in~\eqref{Unione} is connected.
Let us give the following
\begin{Definition}
The \textit{$k$-th open face} of the Minkowski fundamental domain $\mathcal{M}$ is the set of points $\sigma$
belonging to $\mathcal{M}$ such that
\begin{gather*}
f_{k}(\sigma)=0
\qquad
\text{and}
\qquad
f_{j}(\sigma)>0,
\qquad
j\neq k.
\end{gather*}
\end{Definition}

Next lemma provides a~working criterion to individuate some identif\/ications on the boundary of
$\mathcal{M}$ which will serve as ``building blocks'' in our task.
\begin{Lemma}\label{lemma_quarto_cinque}
Let $g$ belong to $\mathbf{GL} (n,\mathbb{Z} )$.
Suppose that $\mathfrak{C}_{g}$ maps one point of one open face of~$\mathcal{M}$ into a~point of an open
face of~$\mathcal{M}$ $($even the same one$)$.
Then the set
$
\text{\rm int} (\mathcal{M}\cup\mathfrak{C} (\mathcal{M},g ) )
$
is connected.
\end{Lemma}

\begin{proof}
Let $\sigma_0$ be the image of $\rho_0$ via some non-identical $\mathfrak{C}_g$.
Assume that these two points belong to the $k_\sigma$-th and to the $k_\rho$-th open face of $\mathcal{M}$
respectively.
There exists an open ball $\mathcal{B}_{\rho_0} \subset {\rm Sym}_{>0} ( n, \mathbb{R}  )$ with
centre in $\rho_0$ such that
\begin{gather*}
f_j (\rho )>0
\qquad\forall\, \rho\in\mathcal{B}_{\rho_0},
\qquad
j\neq k_\rho.
\end{gather*}
Let $\mathcal{B}_{\sigma_0}$ denote the image of $\mathcal{B}_{\rho_0}$ via $\mathfrak{C}_g$.
One can also assume that
\begin{gather*}
f_j (\sigma )>0
\qquad\forall\, \sigma\in\mathcal{B}_{\sigma_0},
\qquad
j\neq k_\sigma.
\end{gather*}
Let us prove that
\begin{gather}
\label{Shan}
\mathcal{B}_{\sigma_0}\subset\mathcal{M}\cup\mathfrak{C}_g (\mathcal{M} ).
\end{gather}
Let $\sigma\in \mathcal{B}_{\sigma_0}\backslash \mathcal{M}$ be the image of $\rho\in\mathcal{B}_{\rho_0}$
via $\mathfrak{C}_g$.
One needs to show that $\rho$ belongs to $\mathcal{M}$, or equivalently, that $f_{k_{\rho}} ( \rho
 )$ is nonnegative.
So, let us suppose that $f_{k_\rho} ( \rho  )$ is strictly negative, instead, and put
\begin{gather*}
\rho=\rho_0+\nu,
\qquad
\sigma=\sigma+\mu.
\end{gather*}
The matrices $\nu$ and $\mu$ are understood to be $ ( n\times n )$-dimensional, symmetric and
dif\/ferent than zero, but not necessarily positive def\/inite.
Let us introduce
\begin{gather*}
\tilde{\rho}=\rho_0-\nu \in \mathcal{B}_{\rho_0},
\qquad
\tilde{\sigma}=\sigma_0-\mu \in \mathcal{B}_{\sigma_0}.
\end{gather*}
By linearity, $\mathfrak{C}_g$ maps $\tilde{\rho}$ into $\tilde{\sigma}$.
On the other side, due to linearity of~$f_{k_\rho}$ and~$f_{k_\sigma}$, both~$f_{k_\rho} ( \tilde{\rho}
 )$ and~$f_{k_\sigma} ( \tilde{\sigma}  )$ are strictly positive.
This implies that both $\tilde{\rho}$ and $\tilde{\sigma}$ belong to $\mathcal{M}$, which is
a~contradiction in view of Lemma~\ref{lem_primo_cinque} and of the assumption that $\mathfrak{C}_g$ is not
the identity.
So $\rho$ belongs to~$\mathcal{M}$ and~\eqref{Shan} holds.
The lemma then follows by means of standard arguments from basic topology.
\end{proof}

Suppose now that the matrices
$
h_{1},h_{2},\ldots,h_{l}\in\mathbf{GL}(n,\mathbb{Z})
$
all satisfy the hypotheses of Lem\-ma~\ref{lemma_quarto_cinque}.
More generally one can consider f\/inite collections of matrices of the form
\begin{gather}\label{famiglia}
{\rm Id},\quad
h_{j_1},\quad h_{j_2}h_{j_1},\quad h_{j_k}h_{j_{k-1}}\cdots h_{j_1},
\end{gather}
where $1\leq j_1,j_2,\ldots,j_k\leq l$.
One can prove that also in this case the set
\begin{gather*}
\text{int}\big(\mathcal{M}\cup\mathfrak{C} (\mathcal{M},h_{j_1} )
\cup\mathfrak{C} (\mathcal{M},h_{j_2}h_{j_1} )\cup\cdots
\cup\mathfrak{C} (\mathcal{M},h_{j_k}h_{j_{k-1}}\cdots h_{j_1} )\big)
\end{gather*}
is connected.
It is possible to use this simple fact to try to determine algorithmically a~set of matrices of
$\mathbf{GL}(n,\mathbb{Z})$ which satisf\/ies the hypotheses of
Proposition~\ref{proposizione_prima_cinque}.

We report an example here below.

\subsection*{The case $\boldsymbol{2g_0=4}$}

Let us denote by
\begin{gather*}
\sigma=\left[
\begin{matrix}
\sigma_{1,1}&\sigma_{1,2}&\sigma_{1,3}&\sigma_{1,4}
\\
\sigma_{1,2}&\sigma_{2,2}&\sigma_{2,3}&\sigma_{2,4}
\\
\sigma_{1,3}&\sigma_{2,3}&\sigma_{3,3}&\sigma_{3,4}
\\
\sigma_{1,4}&\sigma_{2,4}&\sigma_{3,4}&\sigma_{4,4}
\\
\end{matrix}
\right]\in{\rm Sym}_{>0}(4,\mathbb{R})
\end{gather*}
the generic real, symmetric and positive def\/inite matrix of dimension~4.
A fundamental domain~$\mathcal{M}$ for the congruence action of $\mathbf{GL}(4,\mathbb{Z})$ on
${\rm Sym}_{>0}(4,\mathbb{R})$ can be obtained by imposing the following conditions:
\begin{itemize}
\item[i)]
$
\sigma_{1,1} \leq \sigma_{2,2} \leq \sigma_{3,3} \leq \sigma_{4,4}.
$
\item[ii)]
$
\sigma_{1,2}\geq0$,
$\sigma_{2,3}\geq0$,
$\sigma_{3,4}\geq0.
$
\item[iii)]
\begin{gather}\label{Tonio}
\mathbf{m}\cdot\sigma\cdot\mathbf{m}^{T}\geq\sigma_{q,q}
\end{gather}
for every
\begin{gather}\label{def_insieme_m}
\mathbf{m}
\in
\left\{
\begin{matrix}                                                              
(1,0,1,0),&(1,0,0,1),&(0,1,0,1),&(-1,1,0,0),
\\
(-1,0,1,0),&(-1,0,0,1),&(0,-1,1,0),&(0,-1,0,1),
\\
(0,0,-1,1),&(0,1,-1,1),&(1,-1,0,1),&(-1,1,0,1),
\\
(1,0,-1,1),&(-1,0,-1,1),&(1,-1,1,0),&(1,-1,1,1),
\\
(1,1,-1,1),&(-1,-1,1,1),&(-1,1,-1,1),&(1,-1,-1,1)
\end{matrix}
\right\},
\end{gather}
the index $q$ in the right-hand side of~\eqref{Tonio} depending on $\mathbf{m}$ as follows:
\begin{gather*}
q:=\max\big\{j\;\text{such that}\;m_j\neq0,
\;
\mathbf{m}= (m_j )_{j=1}^4\big\}.
\end{gather*}
\end{itemize}
This explicit result is due to E.S.~Barnes and M.J.~Cohn (see~\cite{Barnes-Cohn}).
We used it to determine a~fundamental domain $\mathcal{D}^{\prime\prime}$ for the congruence action of
$\mathbb{K}_{4,1}$ on ${\rm Sym}_{>0}(4,\mathbb{R})$.
Our calculations are summarized here below.
\begin{Observation}
Let $\sigma$ belong to $\mathcal{M}$.
Assume that equality holds in~\eqref{Tonio} for some vector $\mathbf{m}_{0}$ of the
form~\eqref{def_insieme_m}.
Let $\tilde{g}$ be the identity matrix with the $q$-th column replaced by $\mathbf{m}_{0}^{T}$.
This is an upper triangular element of $\mathbf{GL}(4,\mathbb{Z})$.
Using the theory of Minkowski it can be proved that $\mathfrak{C}\left(\sigma,\tilde{g}\right)$ also
satisfies conditions $i)$ and $iii)$ above.
Considering then an appropriate diagonal matrix $d$ belonging to $\mathbf{GL}(4,\mathbb{Z})$ one obtains that
\begin{gather*}
\mathfrak{C}\left(\sigma,\tilde{g}d\right)
\in
\mathcal{M}
\end{gather*}
is equivalent to $\sigma$ and belongs to $\mathcal{M}$.
\end{Observation}

By means of this last observation one can easily determine the following elements of
$\mathbf{GL}(4,\mathbb{Z})$:
\begin{gather*}
h_{1}= \left[
\begin{matrix}
1&0&0&0
\\
0&0&1&0
\\
0&1&0&0
\\
0&0&0&1
\\
\end{matrix}
\right],
\qquad
h_{2}= \left[
\begin{matrix}
0&1&0&0
\\
1&0&0&0
\\
0&0&1&0
\\
0&0&0&1
\\
\end{matrix}
\right],
\qquad
h_{3}= \left[
\begin{matrix}
1&0&1&0
\\
0&1&0&0
\\
0&0&1&0
\\
0&0&0&1
\\
\end{matrix}
\right],
\\
h_{4}= \left[
\begin{matrix}
1&0&0&1
\\
0&1&0&0
\\
0&0&1&0
\\
0&0&0&1
\\
\end{matrix}
\right],
\qquad
h_{5}= \left[
\begin{matrix}
1&0&0&0
\\
0&1&0&1
\\
0&0&1&0
\\
0&0&0&1
\\
\end{matrix}
\right],
\qquad
h_{6}= \left[
\begin{matrix}
-1&-1&0&0
\\
0&1&0&0
\\
0&0&1&0
\\
0&0&0&1
\\
\end{matrix}
\right],
\\
h_{7}= \left[
\begin{matrix}
1&0&0&0
\\
0&1&1&0
\\
0&0&-1&0
\\
0&0&0&-1
\\
\end{matrix}
\right],
\qquad
h_{8}= \left[
\begin{matrix}
1&0&0&0
\\
0&1&0&0
\\
0&0&1&1
\\
0&0&0&-1
\\
\end{matrix}
\right],
\qquad
h_{9}= \left[
\begin{matrix}
1&0&0&0
\\
0&1&0&-1
\\
0&0&1&1
\\
0&0&0&-1
\\
\end{matrix}
\right],
\\
h_{10}= \left[
\begin{matrix}
1&0&0&1
\\
0&1&0&-1
\\
0&0&1&0
\\
0&0&0&1
\\
\end{matrix}
\right],
\qquad
h_{11}= \left[
\begin{matrix}
1&0&0&-1
\\
0&1&0&0
\\
0&0&1&1
\\
0&0&0&-1
\\
\end{matrix}
\right],
\qquad
h_{12}= \left[
\begin{matrix}
1&0&-1&0
\\
0&1&1&0
\\
0&0&-1&0
\\
0&0&0&-1
\\
\end{matrix}
\right].
\end{gather*}
They all satisfy the hypotheses of Lemma~\ref{lemma_quarto_cinque}. This can be verif\/ied considering for each matrix above the corresponding point of $\mathcal{M}$ from
the list below:
\begin{gather}
\rho_{0}^{1}= \left[
\begin{matrix}
10&1&1&1
\\
1&11&1&1
\\
1&1&11&1
\\
1&1&1&12
\\
\end{matrix}
\right],\qquad\rho_{0}^{2}= \left[
\begin{matrix}
10&1&1&1
\\
1&10&1&1
\\
1&1&11&1
\\
1&1&1&12
\\
\end{matrix}
\right],\qquad\rho_{0}^{3}= \left[
\begin{matrix}
10&1&-5&1
\\
1&11&1&1
\\
-5&1&12&1
\\
1&1&1&13
\\
\end{matrix}
\right],
\nonumber
\\
\rho_{0}^{4}= \left[
\begin{matrix}
10&1&1&-5
\\
1&11&1&1
\\
1&1&12&1
\\
-5&1&1&13
\\
\end{matrix}
\right],\qquad\rho_{0}^{5}= \left[
\begin{matrix}
11&1&1&1
\\
1&12&1&-6
\\
1&1&13&1
\\
1&-6&1&14
\\
\end{matrix}
\right],\qquad\rho_{0}^{6}= \left[
\begin{matrix}
10&5&-1&1
\\
5&11&1&1
\\
-1&1&12&1
\\
1&1&1&13
\\
\end{matrix}
\right],
\nonumber
\\
\rho_{0}^{7}= \left[
\begin{matrix}
11&1&1&1
\\
1&12&6&-1
\\
1&6&13&1
\\
1&-1&1&14
\\
\end{matrix}
\right],\qquad\rho_{0}^{8}= \left[
\begin{matrix}
10&1&1&1
\\
1&11&1&1
\\
1&1&12&6
\\
1&1&6&13
\\
\end{matrix}
\right],\qquad\rho_{0}^{9}= \left[
\begin{matrix}
4&1&1&1
\\
1&5&1&-2
\\
1&1&7&3
\\
1&-2&3&8
\\
\end{matrix}
\right],
\nonumber
\\
\rho_{0}^{10}= \left[
\begin{matrix}
3&1&1&-1
\\
1&5&1&2
\\
1&1&6&1
\\
-1&2&1&7
\\
\end{matrix}
\right],\!\qquad\rho_{0}^{11}= \left[
\begin{matrix}
3&1&1&-1
\\
1&4&1&1
\\
1&1&5&2
\\
-1&1&2&6
\\
\end{matrix}
\right],\!\qquad\rho_{0}^{12}= \left[
\begin{matrix}
3&1&-1&1
\\
1&5&2&1
\\
-1&2&6&1
\\
1&1&1&7
\\
\end{matrix}
\right].\!\!\!\!
\label{fondamentali}
\end{gather}

A set of elements of $\mathbf{GL}(4,\mathbb{Z})$ satisfying the hypotheses of
Proposition~\ref{proposizione_prima_cinque} is the following:
\begin{alignat}{4}
&g_{1}={\rm Id},\qquad&
&g_{11}=h_{7}h_{5}h_{1},\qquad&
&g_{21}=h_{9}h_{2},&
\nonumber
\\
&g_{2}=h_{1},\qquad&
&g_{12}=h_{12}h_{5}h_{1},\qquad&
&g_{22}=h_{4},&
\nonumber
\\
&g_{3}=h_{3}h_{1},\qquad&
&g_{13}=h_{7}h_{1},\qquad&
&g_{23}=h_{6}h_{4},&
\nonumber
\\
&g_{4}=h_{4}h_{3}h_{1},\qquad&
&g_{14}=h_{10}h_{7}h_{1},\qquad&
&g_{24}=h_{6},&
\nonumber
\\
&g_{5}=h_{5}h_{3}h_{1},\qquad&
&g_{15}=h_{10}h_{1},\qquad&
&g_{25}=h_{2}h_{6},&
\nonumber
\\
&g_{6}=h_{9}h_{3}h_{1},\qquad&
&g_{16}=h_{12}h_{10}h_{1},\qquad&
&g_{26}=h_{9}h_{6},&
\nonumber
\\
&g_{7}=h_{4}h_{1},\qquad&
&g_{17}=h_{12}h_{1},\qquad&
&g_{27}=h_{8},&
\nonumber
\\
&g_{8}=h_{7}h_{4}h_{1},\qquad&
&g_{18}=h_{2},\qquad&
&g_{28}=h_{11}.&
\nonumber
\\
&g_{9}=h_{12}h_{4}h_{1},\qquad&
&g_{19}=h_{3}h_{2},\qquad&
&&
\nonumber
\\
&g_{10}=h_{5}h_{1},\qquad&
&g_{20}=h_{6}h_{2},\qquad&
&&
\label{tutti_definitivi}
\end{alignat}

Indeed, by means of a~computer it is easy to verify that no two such matrices belong to the same left coset
and, in view of~\eqref{indice_esplicito}, the index of $\mathbb{K}_{4,1}$ in
$\mathbf{GL}(4,\mathbb{Z})$ is 28; so this list is composed of exactly one representative for
each left coset.
Moreover,~\eqref{tutti_definitivi} is the union of families of the type~\eqref{famiglia}; thus it gives
place to a~set whose interior is connected.

We emphasize our f\/inal result in the following
\begin{Theorem}
A fundamental domain $\mathcal{D}^{\prime\prime}$ for the congruence action of $\mathbb{K}_{4,1}$ on ${\rm
Sym}_{>0}(4,\mathbb{R})$ is given by
\begin{gather*}
\mathcal{D}^{\prime\prime}=\bigcup_{j=1}^{28}\mathfrak{C}(\mathcal{M},g_j),
\end{gather*}
where the matrices $g_{j}$ are listed in~\eqref{tutti_definitivi}.
\end{Theorem}

\section{Discussion}

The method presented here works, in principle, when $2g_0$ is an arbitrary positive and even integer but it
requires the explicit description of the Minkowski fundamental domain $\mathcal{M}$.
Such a~description, though not completely non-redundant, is available for $2g_0=6$ (see~\cite{Ryshkov}); it
would be interesting to try to work out the calculations in this case, with a~more systematic use of
a~computer.
To the best of our knowledge, no explicit description of $\mathcal{M}$ is yet available when $2g_0$ is
equal or larger than 8.

As already mentioned in the Introduction, the problem of f\/inding a~canonical form for period matrices of
real Riemann surfaces was not f\/irst studied in this article.
In~\cite{Silhol}, Silhol proposed a~dif\/ferent approach, in which the original problem can be reduced to
the Minkowski problem itself.
No restrictions on the group $\mathbf{GL}(n,\mathbb{Z})$ are necessary while considering its
congruence action on ${\rm Sym}_{>0}(n,\mathbb{R})$.
On the other side, the locus of canonical forms for period matrices of topological type $(2g_0, 1,0)$, as
described in~\cite{Silhol}, is not a~connected set for $2g_0\geq 4$.
Our fundamental domain is, instead (as required in Def\/inition~\ref{def_dom_fond}).
Moreover, in our approach the bases in the homology under consideration all behave in the same way w.r.t.\
$\mathbf{r}_{\star}$ (this constraint is expressed by formulas~\eqref{Intro_3}).
As a~consequence, geometrical quantities like periods of abelian dif\/ferentials exhibit some specif\/ic
symmetries when computed w.r.t.\
these basis.
This turns out to be a~good advantage in the application of our results to the study of algebro-geometric
solutions to nonlinear integrable systems.

\subsection*{Acknowledgements} Research supported by SISSA under the PhD program in Mathematics and by the
Austrian Science Fund (FWF) under Grant No.~Y330.
The author wishes to thank Professor Boris Dubrovin for kindly supervising this work and Professor Tamara
Grava for valuable discussions.
He also thanks the anonymous referees for signif\/icantly contributing to improve this article.

\pdfbookmark[1]{References}{ref}
\LastPageEnding

\end{document}